\documentclass[11pt,a4paper,english]{article}
\usepackage{graphicx}
\usepackage{amssymb,latexsym}
\usepackage{amsmath}
\usepackage{epsf}
\usepackage{pdfsync}
\usepackage{epsfig}
\usepackage[parfill]{parskip}
\usepackage[matrix,arrow,color]{xy}
\usepackage[usenames]{color}
\usepackage[bookmarks=false]{hyperref}
\usepackage{mathrsfs}
\usepackage[mathcal]{euscript}
\usepackage[font={small}]{caption}
\usepackage{subcaption}
\usepackage[export]{adjustbox}
\usepackage{float}
\usepackage{wrapfig}
\usepackage{microtype}
\usepackage{slashed}
\usepackage [latin1]{inputenc}

\input{xy}
\xyoption{all}

\input{epsf}

\makeatletter

\usepackage{setspace}

\usepackage{lscape}

\catcode`@=11
\renewcommand{\section}
{\@startsection{section}{1}{0pt}{\medskipamount}{\medskipamount}{\large\bf}}
\makeatletter\renewcommand{\subsection}
{\@startsection{subsection}{2}{\z@}{-3.25ex plus -1ex minus -.2ex}
{1.5ex plus .2ex}{\it }}

\numberwithin{equation}{section}
\catcode`@=12

\newcommand{\ban}{\begin{eqnarray}}
\newcommand{\ean}{\end{eqnarray}}

\newcommand{\Tr}{{\rm Tr}}

\newcommand{\cW}{{\cal W}}
\newcommand{\cN}{{\cal N}}
\newcommand{\cM}{{\cal M}}
\newcommand{\cS}{{\cal S}}
\newcommand{\cB}{{\cal B}}

\newcommand{\cH}{{\cal H}}

\newcommand{\cO}{{\cal O}}
\newcommand{\cC}{{\cal C}}

\newcommand{\sfX}{{\mathsf{X}}}

\newcommand{\DT}{{\tt DT}}

\newcommand{\Hecke}{\mathcal{H\kern-.77em H}}

\newcommand{\mbf}[1]{{\boldsymbol {#1} }}
\newcommand{\complex}{{\mathbb C}} 
\newcommand{\zed}{{\mathbb Z}} 
\newcommand{\real}{{\mathbb R}} 

\def\e{{\,\rm e}\,}

\def\ii{{\,{\rm i}\,}}

\newcommand{\sign}{\mathrm{sgn}}

\def\beq{\begin{equation}}
\def\bee{\begin{equation}}
\def\eeq{\end{equation}}
\def\bea{\begin{eqnarray}}
\def\eea{\end{eqnarray}}
\def\bd{\begin{displaymath}}
\def\ed{\end{displaymath}}

\newcommand{\Cint}{\int\kern-10.5pt-\kern7pt}

\newcommand{\PP}{{\mathbb{P}}}

\newcommand{\be}{\begin{equation}}
\newcommand{\ee}{\end{equation}}
\newcommand{\bal}{\begin{align}}
\newcommand{\eal}{\end{align}}

\newcommand\fverbit{\egroup\item[\fbox{\unhbox\pippobox}]}
\newbox\pippobox

{

\def\be{\begin{equation}}
\def\ee{\end{equation}}
\def\bea{\begin{eqnarray}}
\def\eea{\end{eqnarray}}

\begin{document}

\begin{titlepage}
\setcounter{page}{1}

\vskip 5cm

\begin{center}

\vspace*{3cm}

{\Huge A note on discrete dynamical systems \\[8pt ]in theories of class $S$}

\vspace{15mm}

{\large\bf Michele Cirafici}
\\[6mm]
\noindent{\em Dipartimento di Matematica e Geoscienze, Universit\`a di Trieste, \\ Via A. Valerio 12/1, I-34127 Trieste, Italy, 
\\ Institute for Geometry and Physics (IGAP), via Beirut 2/1, 34151, Trieste, Italy
\\ INFN, Sezione di Trieste, Trieste, Italy 
}\\[4pt] Email: \ {\tt michelecirafici@gmail.com}

\vspace{15mm}

\begin{abstract}
\noindent

In this note we consider the set of line operators in theories of class $S$. We show that this set carries the action of a natural discrete dynamical system associated with the BPS spectrum. We discuss several applications of this perspective; the relation with global properties of the theory; the set of constraints imposed on the spectrum generator, in particular for the case of SU(2) $\cN=2^*$; and the relation between line defects and certain spherical Double Affine Hecke Algebras.

\end{abstract}

\vspace{15mm}

\today

\end{center}
\end{titlepage}

\newpage

\tableofcontents


\section{Introduction and discussion}







In this note we consider $\cN=2$ supersymmetric quantum field theories of class $\cS$ and construct a discrete dynamical system which acts on their set of line operators. The purpose is to introduce techniques from dynamical systems to study the topology of such a set and discuss their physical implications.

Theories of class $\cS$ can be engineered by compactifying the $\cN=(2,0)$ superconformal theory in six dimensions on a curve $\cC$. As a consequence their properties can be studied using geometrical tools associated with $\cC$, such as spectral networks \cite{Gaiotto:2009hg,Gaiotto:2012rg}, or algebraic tools, such as quivers \cite{Alim:2011kw}.

These theories admit supersymmetric line defects which are associated to the spectral problem of computing framed BPS states \cite{Gaiotto:2010be}. In many cases such a problem can be approached by localization on quiver moduli spaces \cite{Chuang:2013wt,Cirafici:2017wlw,Cirafici:2018jor,Cirafici:2019otj,Cordova:2013bza}, or on the moduli spaces of semiclassical configurations \cite{Brennan:2018yuj,Moore:2015szp}, by using cluster algebras \cite{Allegretti:2018arr,Cirafici:2013bha,Cirafici:2017iju,Williams:2014efa}, or via BPS graphs or spectral networks \cite{Gabella:2016zxu,Gabella:2017hpz,Gang:2017ojg,Neitzke:2020jik}.

In this note we revisit and extend the approach of \cite{Cirafici:2013bha,Cirafici:2017iju} taking inspiration from the perspective advocated in \cite{Dimitrov:2013xpa}. In quiver language the BPS spectral problem can be solved by looking for sequences of quiver mutations which scan over all the physical BPS states. Such sequences are associated with chambers with finite number of BPS particles and have the property that they take the BPS quiver back to itself, eventually up to a relabelling of the nodes. To such a sequence one can associate a rational transformation, sometimes called the spectrum generator \cite{Gaiotto:2009hg}, the (quantum) half-monodromy \cite{Cecotti:2010fi}, or the Kontsevich-Soibelman transformation \cite{Kontsevich:2008fj}, up to a certain freedom in the conventions. Different ways of decomposing this transformation encode the BPS spectrum in every chamber of the theory.

A line operator can be described in terms of a framed BPS quiver. Such sequences of mutations also act on framed BPS quivers, in general mapping a framing to another framing (not necessarily different). The above rational transformations, and the invariance of the line operator under wall-crossing, induce an action on the  set of line operators \cite{Cirafici:2013bha}. This action has the form of a rational transformation which applied to the vev of the line operator, viewed as a sum of monomials in the cluster variables whose coefficients are the framed BPS degeneracies, produces the vev of a different line operator. The iteration of this operation produces a family of line operators, of which all the vevs can be computed explicitly \cite{Cirafici:2013bha}. Under certain conditions this iteration is related to certain integrable systems \cite{Cirafici:2017iju}, although in this note we will be more general.

In this note we shift slightly the perspective: we use this operation to define an abstract dynamical system $\mathsf{R}$ which acts on the set of line operators $\mathsf{Line}$. Abstract (discrete) dynamical systems are iterated maps acting on a space and can be used to study the properties of such a space, for example in the topological or geometrical sense. The idea is then to use the properties of $\mathsf{R}$ to study the set $\mathsf{Line}$: for example $\mathsf{R}$ will have fixed points, closed cycles or infinite orbits, and from this information we might hope to gain knowledge about the structure of $\mathsf{Line}$ as a topological space. 

This approach is partly inspired by the relation between supersymmetric quantum mechanics and Morse theory \cite{Witten:1982im}, where one studies a supersymmetric particle propagating in a compact Riemannian manifold. In that case the central role is played by a continuous dynamical system, the instanton equations. Enumerative invariants constructed by counting solutions which connect fixed points determine the cohomology of the Morse-Smale-Witten complex and therefore contain geometric information about the underlying manifold. While our construction is quite different, this is a useful analogy to keep in mind.

After collecting some results about theories of class $S$ and line defects in Section \ref{classS}, we discuss our dynamical systems in Section \ref{DynSys}. 
In particular in this note we discuss three applications:
\begin{itemize}
\item In the UV the labels of line operators can be used to distinguish different quantum field theories \cite{Aharony:2013hda}. In Section \ref{reading} we show with an explicit example that in the IR this labelling is encoded in the orbits of the dynamical system $\mathsf{R}$.
\item The condition that certain line operators are fixed points of the discrete dynamical system is sometimes strong enough to compute the spectrum generator as a rational transformation. We show this in Section \ref{fixedpoints} in the cases of pure SU(2) and SU(2) $\cN=2^*$ theories.
\item The quantum dynamical system can be used to study the algebraic relations between line operators. As an example in Section \ref{sDAHA} we show that certain fixed points give a presentation of the spherical Double Affine Hecke Algebra in the cases of $A_1$ e  $C^\vee C_1$.
\end{itemize}
Since the details are somewhat technical, as they involve repeated iterations of complicated rational transformations, a supporting \textsc{mathematica} file which includes the computations of Sections 4--6 is made available at \cite{mathe}.

This perspective naturally leads to a series of questions. The most natural is if other tools taken from the theory of dynamical systems, such as  invariant sets or Conley theory, have natural applications in the framework  of theories of class $S$. Furthermore, while in this paper we present our results in purely algebraic terms, using quivers, there should be a geometrical counterpart in terms of data on the curve $\cC$. In particular it would be interesting to understand more clearly how the action on representation theory objects is induced by geometrical quantities, possibly at the categorical level along the lines of \cite{GTW}. Another interesting problem is the relation between our approach here and certain algebraic structures which appear in the UV description \cite{Radler:2018pvl}. Other approaches using discrete dynamical system in the framework of quivers in supersymmetric theories include \cite{Bonelli:2020dcp,Cecotti:2014zga,Neitzke:2017cxz}.

\section{Theories of class $S$ and their line defects} \label{classS}

In this Section we will briefly introduce certain aspects of theories of class $\cS$ which we will need in the rest of the note: the geometry of the Hitchin moduli spaces, the rational trasformations associated with quivers and certain properties of line operators.

\subsection{Theories of class $S$ and Hitchin systems}


Theories of class $S$ are four dimensional quantum field theories which can be obtained by compactifying the six dimensional $\cN=(2,0)$ theory on a curve $\cC$, with punctures and additional data specified at the punctures. Such theories have extended supersymmetry and moduli spaces of quantum vacua. When the theory is defined on $\real^4$ at a generic point of the Coulomb branch $\cB$ the gauge symmetry is broken down to the maximal torus of the gauge group $G$. When the theory is further compactified down to  $\real^3 \times S^1_R$ the moduli space of vacua $\cM_H$ is identified with the Hitchin moduli space associated with $\cC$. The Hitchin moduli space $\cM_H$ is hyperK\"ahler: it carries a family of complex structures $J_\zeta$ parametrized by $\zeta \in \PP^1$, with holomorphic symplectic forms $\omega_\zeta$. In complex structure $J_0$ it can be locally seen as a fibration over $\cB$, whose fibers are compact tori.

From this perspective the Seiberg-Witten curve $\Sigma_u$ arises as the spectral curve associated with the Hitchin fibration. The Seiberg-Witten solution determines the low energy Wilsonian action on the Coulomb branch $\cB$ in terms of $(\Sigma_u , \lambda_u)$. In particular the lattice of electric and magnetic charges of the theory is identified as $\Gamma = H_1 (\Sigma , \zed)$ and the central charge operator becomes a function $Z (u) \, : \, \Gamma \longrightarrow \complex$ which depends holomorphically on $u \in \cB$ and is completely determined by the periods of the Seiberg-Witten differential $\lambda_u$.

The geometry of the moduli space $\cM_H$ can be expressed in terms of a set of Darboux coordinates $\{ X_{\gamma} (u , \zeta ) \}$ \cite{Gaiotto:2009hg}. Such coordinates are associated with stable BPS particles of charge $\gamma \in \Gamma$. They obey the Poisson bracket
\begin{equation}
\{ X_{\gamma}  ,  X_{\gamma'} \} = \langle \gamma , \gamma^\prime \rangle \, X_{\gamma + \gamma'} \, ,
\end{equation}
and satisfy the twisted group algebra
\begin{equation}
X_{\gamma} \, X_{\gamma'} = (-1)^{\langle \gamma , \gamma' \rangle} \ X_{\gamma+ \gamma'} \, .
\end{equation}
The coordinates $\{ X_{\gamma} (u , \zeta ) \}$ jump at real codimension one walls in $\mathcal{B} \times \mathbb{C}^*$ associated with BPS particles, precisely at the loci where
\begin{equation}  \label{BPSray}
Z(\gamma')/\zeta \in \mathbb{R}_- \, ,
\end{equation}
which are called BPS rays. The discontinuity is given by the rational transformation
\begin{equation} \label{jumpX}
X_{\gamma} \longrightarrow \mathcal{K}_{\gamma'}^{\Omega (\gamma' , u)} (X_{\gamma})
\end{equation}
expressed in terms of the Kontsevich-Soibelman symplectomorphism \cite{Kontsevich:2008fj}
\begin{equation}
\mathcal{K}_{\gamma'} (X_{\gamma}) = X_{\gamma} \, \left( 1 - X_{\gamma'} \right)^{\langle \gamma, \gamma' \rangle} \ ,
\end{equation}
and of the degeneracy of BPS states $\Omega (\gamma' , u)$.

It will be useful in the following to introduce untwisted coordinates $Y_{\gamma}$, so that
\begin{equation}
Y_{\gamma} \, Y_{\gamma'} = Y_{\gamma+ \gamma'} \ .
\end{equation}
These coordinates describe locally a space $\widetilde{\mathcal{M}}_H$ and up to a quadratic refinement can be identified with the Darboux coordinates $X$ on $\mathcal{M}_H$, establishing an isomorphism between $\mathcal{M}_H$ and $\widetilde{\mathcal{M}}_H$ \cite{Gaiotto:2010be}. The isomorphism between the two sets of coordinates is given by a quadratic refinement, a map $ \sigma \, : \, \Gamma \longrightarrow \{ \pm 1 \}$ such that
\begin{equation}
\sigma (\gamma) \sigma (\gamma') = (-1)^{\langle \gamma , \gamma' \rangle} \, \sigma (\gamma + \gamma') \, .
\end{equation}
In the following we will identify $Y_\gamma$ with $\sigma (\gamma) X_\gamma$ with the choice $\sigma (\gamma) = (-1)^{J_3 + I_3}$ \cite{Gaiotto:2010be}. With these choices, the transformation law for the coordinates $Y_\gamma$ upon crossing the BPS wall for a single hypermultiplet with charge $\gamma'$ becomes a cluster transformation
\begin{equation}
Y_\gamma \longrightarrow Y_\gamma (1 + Y_{\gamma'})^{\langle \gamma , \gamma' \rangle} \ .
\end{equation}
Such a transformation endows $\mathcal{M}_H$ locally with the structure of a cluster variety. In this note we will mostly use the coordinates $Y_\gamma$.

%

Conjecturally such coordinates can be quantized \cite{Cecotti:2010fi,Gaiotto:2010be} in terms of noncommuting operators
$\mathsf{X}_\gamma ( u , \zeta)$ which obey the quantum torus algebra $\mathbb{T}_\Gamma$ associated with the lattice of charges $\Gamma$
\be
\sfX_\gamma \, \sfX_{\gamma'} \, q^{-\langle \gamma , \gamma' \rangle} =  \sfX_{\gamma + \gamma'} \, .
\ee
This way of writing the operators $\mathsf{X}_\gamma$  is also known as \textit{normal ordering} \cite{Cecotti:2010fi}. The coordinates  $\mathsf{X}_\gamma ( u , \zeta)$ jump across BPS walls as their classical limit. The transformation is given by a quantum version of the Kontsevich-Soibelman diffeomorphism, as we will see momentarily.

\subsection{Quivers and quantum monodromies}

In this paper we will only consider theories of quiver type  \cite{Alim:2011kw,Cecotti:2010fi}, where the BPS spectrum can be understood via representation theory methods. Concretely this means that we can find a positive basis $\{ e_i \}$ of $\Gamma$ such that all basis elements correspond to stable hypermultiplets and such that the central charges $Z_{e_i} (p)$ take values in $\mathfrak{h}_\theta  \equiv e^{-i\theta} \mathfrak{h}$ where $\theta \in S^1$ and $\mathfrak{h}$ is the upper half plane. In particular the charge of any BPS state can be expressed as $\gamma = \sum_{i} d_i \, e_i$ where all the $d_i$'s are positive integers.

To such a basis we associate a quiver $Q$ where the set of nodes $Q_0$ is in bijection with the basis elements $\{ e_i \}$, and there are $B_{ij} = \langle e_i ,\, e_j \rangle$ arrows directed from the node $i \in Q_0$ to the node $j \in Q_0$ whenever $\langle e_i , e_j \rangle > 0$. When a theory is of quiver type, the low energy dynamics of a particle is described by an effective quantum mechanics with four supercharges based on the quiver, interacting via a superpotential $\cW$. Stable BPS particles correspond to supersymmetric ground states. Mathematically a BPS  state with charge $\gamma = \sum_i d_i \, e_i$, with $d_i \in \mathbb{Z}_+$, corresponds to a stable representation, where the $d_i$ are the dimensions of the representation spaces. The stability condition is induced by the central charge function $Z_{\gamma} (u)$.

Two quivers obtained from a change of basis describe the same physics. If we require the two basis to obey the conditions stated before, the corresponding quivers are related by a quiver mutation  \cite{Alim:2011kw,Cecotti:2010fi}(up to trivial transformations, such as permutations of the nodes). Physically an elementary quiver mutation correspond to the situation where the central charge of a basis element $e_k$ gets rotated out of the upper half plane $\mathfrak{h}_\theta$, for example by tuning the physical parameters. This happens precisely at the BPS rays \eqref{BPSray}. As the vector corresponding to $e_k$ exits $\mathfrak{h}_\theta$, the vector corresponding $- e_k$ enters $\mathfrak{h}_\theta$.  The mutation operators $\mu^\pm_{e_k}$ act as \cite{FG,FZ}
\be \label{mutpm}
\hat{e}_i \equiv \mu^{\pm}_{e_k}(e_i) \equiv \begin{cases} -e_i &\text{if } i = k \\ e_i + \text{max}(0,\pm\langle e_k ,\, e_i \rangle) e_k &\text{otherwise.}\end{cases}
\ee
The $\pm$ sign above is chosen according to the fact that $\hat{e}$ exits the upper half plane $\mathfrak{h}_\theta$ along $e^{-i\theta}\mathbb{R}_{\mp}$. One can see that $\mu_{\hat{e}}^+ \mu_{\hat{e}}^- = \text{id} = \mu_{\hat{e}}^- \mu_{\hat{e}}^+$. From the definition of the adjacency matrix of the quiver $B_{ij}$, that both transformations $\mu^\pm$ in \eqref{mutpm} correspond to
\begin{equation} \label{mutB}
B'_{ij} = \left\{ \begin{matrix} - B_{ij} & \text{if} \ i=k  \ \text{or} \ j=k \\ B_{ij} + \text{sgn} (B_{ik}) [ B_{ik} \, B_{kj} ]_+ & \text{otherwise} \end{matrix} \right. \ .
\end{equation}

Of particular importance are the sequence of mutations corresponding to a chamber with a finite BPS spectrum. Since we have a finite spectrum we can rotate $\mathfrak{h}_\theta$ clockwise until we reach $\mathfrak{h}_{\theta+\pi}$. Every time $\mathfrak{h}_{\theta}$ crosses a BPS ray, the positive basis changes by an elementary quiver mutation. After crossing all the BPS rays, the basis of charges associated with the half-plane $\mathfrak{h}_{\theta+\pi}$ is the \textsc{cpt} conjugate of the basis we started with. Therefore any chamber with a finite spectrum is associated to a finite sequence $\mathbf{m}^+$ of $N$ of elementary mutations at nodes $i_1, i_2,i_3,...,i_N \in Q_0$. In particular after such a sequence of mutation the quiver is back to itself, eventually up to a permutation: $\mathbf{m}^+(Q) = \pi(Q)$ with $\pi$ a permutation of the labels of the nodes of $Q$.

%

This construction can be refined whenever the spectrum has a $\mathbb{Z}_{2s}$ discrete symmetry, induced by the action of a discrete R-symmetry which acts on the central charge operator as $Z \to e^{i\pi/s} Z$. This implies that the BPS rays are distributed in a $\mathbb{Z}_{2s}$ symmetric manner in the $\mathfrak{h}_\theta$ half-plane. In this case the mutation operator $\mathbf{m}^+$ can be written as the iteration of a smaller sequence of mutations $\mathbf{r}^+$ corresponding to a single $\mathbb{Z}_{2s}$ sector. 

It will be important for our construction that quiver mutations, defined algebraically on a basis of charges $\{ e_i \}$, lift to cluster transformations on a set of coordinates $\{ Y_{e_i} \}$ associated with the same basis, and to quantum cluster transformations on the coordinates $\{ \sfX_{e_i} \}$. We define the quantum mutation $\mathcal{Q}_i$ at a node $i \in Q_0$ to be the composition of a quiver mutation at the level of the charges and the automorphism given by conjugation \cite{FG2}
\begin{equation}
\mathcal{Q}_k \circ \mathsf{Y}_{\gamma_i} = \mathrm{Ad}' \left( \Psi (-\mathsf{Y}_{\gamma_k} ; q) \right) \circ \mu_k^+ (\mathsf{Y}_{\gamma_i}) \equiv
 \Psi (-\mathsf{Y}_{\gamma_k} ; q)^{-1} \, \mathsf{Y}_{\mu_k^+ (\gamma_i)} \Psi (- \mathsf{Y}_{\gamma_k} ; q) \, .
\end{equation}
Similarly one can define an operator $\tilde{\mathcal{Q}}_k$ using $\mu^-_k$.

Consider now a sequence of mutations $\mathbf{m}^+$ associated with the stable BPS spectrum in a certain chamber. Then the corresponding sequence of quantum mutations
\begin{equation} \label{QseqHalf}
\mathcal{Q}_{\mathbf{m}^+} = \mathrm{Ad}' \left( \mathscr{I}_{\pi^{-1}} \circ \mathscr{H}_q \right)
\end{equation}
gives the adjoint action of the half-monodromy operator $ \mathscr{H}_q $, up to the change of basis in $\Gamma$ corresponding to a permutation. The operator $ \mathscr{I}_{\pi^{-1}}$ encodes the lift of this permutation to the quantum torus $\mathbb{T}_\Gamma$. These ideas extend to the case where the theory admits a smaller fractional monodromy. Assume that the $1/m$ fractional monodromy corresponds to the sequence of mutations $\mathbf{r}^+$ and denote by $\sigma$ the associated permutation. Then the analog statement to \eqref{QseqHalf} holds: the sequence of quantum monodromies
\begin{equation} \label{QseqFrac}
\mathcal{Q}_{\mathbf{r}^+} = \mathrm{Ad}' \left( \mathscr{I}_{\sigma^{-1}} \circ \mathscr{Y}_q \right)
\end{equation}
computes the adjoint action of the $1/m$ quantum fractional monodromy, up to a permutation of the basis of $\mathbb{T}_\Gamma$. 
The full quantum monodromy $\mathscr{M}_q = \mathscr{Y}_q^m$ is a wall-crossing invariant and contains the full information about the spectrum of BPS states in every chamber of the theory. The equivalence of different decompositions of $\mathscr{M}_q$ in different chambers is the KS wall-crossing formula.



We will be particularly interested in the action of the quantum monodromy operators on the cluster-like coordinates on the Hitchin moduli space $\mathcal{M}_H$. In the $q \longrightarrow +1$ limit, the quantum mutation $\mathcal{Q}_k$ reduces to a cluster transformation \cite{FG,FG2,FZ}
\begin{equation} \label{clusterR}
\mathcal{Q}_k \cdot Y_i \bigg|_{q \rightarrow +1} \equiv R_i^{(k)} [\{ Y_1 , \dots , Y_n \} ]=  \left\{ 
\begin{array}{cc}
Y_i^{-1} & \text{if} \ i = k \\
Y_i \, \left( 1 + Y_k^{- \sign \langle \gamma_k , \gamma_i \rangle} \right)^{- \langle \gamma_k , \gamma_i \rangle} & \text{if} \ i \neq k 
\end{array}
\right. \, ,
\end{equation}
and similarly
\begin{equation} \label{clusterRt}
\mathcal{\tilde{Q}}_k \cdot Y_i \bigg|_{q \rightarrow +1} \equiv \tilde{R}_i^{(k)} [\{ Y_1 , \dots , Y_n \} ] =  \left\{ 
\begin{array}{cc}
Y_i^{-1} & \text{if} \ i = k \\
Y_i \, \left( 1 + Y_k^{ \sign \langle \gamma_k , \gamma_i \rangle} \right)^{\langle \gamma_k , \gamma_i \rangle} & \text{if} \ i \neq k 
\end{array}
\right. \, .
\end{equation}
Here we have introduced the notation $Y_i = Y_{e_i}$. Note that the rational transformation $\tilde{R}_i^{(k)}$ in \eqref{clusterRt} coincides with the transformation $R_i^{(k)}$ in \eqref{clusterR} for the opposite quiver. 

In particular we can consider rational transformations arising from the $q \longrightarrow +1$ limit of the half monodromy or of any fractional monodromy. Let $\mathbf{r}^+ = i_1 , i_2 , \cdots i_k$ be a sequence of quiver mutations corresponding to the finest fractional monodromy (eventually including the half-monodromy), within a finite chamber in a $\mathcal{N}=2$ model. Let $\sigma$ be the associated permutation. Then we define the rational transformations
\begin{align}  \label{RonCluster}
R_j [ \{ Y_1 , \dots , Y_n \} ] &= \sigma^{-1} R_j^{(i_k)} \, R_j^{(i_{k-1})} \, \cdots R_j^{(i_1)} [\{ Y_1 , \cdots , Y_n \}] \, , \\ \label{RtonCluster}
\tilde{R}_j [ \{ Y_1 , \dots , Y_n \} ] &= \sigma^{-1}  \tilde{R}_j^{(i_k)} \, \tilde{R}_j^{(i_{k-1})} \, \cdots \tilde{R}_j^{(i_1)} [\{ Y_1 , \cdots , Y_n \}] \, .
\end{align}
The recursive relation
\begin{equation}
Y_{j , s+1} =  R_j [ \{ Y_{1,s} , \dots , Y_{n,s} \} ] 
\end{equation}
(and similarly for $\tilde{R}$), defines a \textit{discrete dynamical system} associated to the finite chamber $\mathscr{C}$ which in certain cases can be shown to be integrable \cite{Cirafici:2017iju}.  Such a dynamical system acts on the cluster coordinates associated to the BPS quiver. After an evolution given by one unit of time, the underlying quiver is back to itself (as a graph), since we are taking into account the permutation $\sigma$, while all the cluster variables have evolved by a sequence of mutations. The transformation $R_j$ expresses the result of the action of the sequence of cluster transformations on the $j$-th cluster variable $Y_j$. We will denote by $R_{\mathbf{r}^+}$ or $R_{\mathbf{m}^+}$ the  operation which computes the action of the rational transformations associated with sequences of mutations $\mathbf{r}^+$ and $\mathbf{m}^+$ on all the cluster variables. Similarly we introduce the transformations $\tilde{R}_{\mathbf{m}^+} $ and $\tilde{R}_{\mathbf{r}^+} $.

In this note we will consider a different dynamical system, given by the lift of the $R$ and $\tilde{R}$ transformations to the space of line operators.
 
%
%

\subsection{Line defects and framed BPS quivers}

Theories of class $S$ admit BPS line defects, described by straight lines in $\mathbb{R}^{1,3}$ extended in the time direction. When a theory has a lagrangian description based on a Lie algebra $\mathfrak{g}$ such line defects admit a UV labelling by pairs of weights $ \alpha = (\lambda_e , \lambda_m) \in \Lambda_w \times \Lambda_{mw} / \mathfrak{w}$, which lie in the weight lattice of $\mathfrak{g}$ and the weight lattice of the Langlands dual algebra $\mathfrak{g}^*$ respectively, and are considered modulo the action of the Weyl group $\mathfrak{w}$. They are further constrained by the quantization condition
\begin{equation}
\langle (\lambda_e , \lambda_m) , (\lambda'_e , \lambda'_m) \rangle \in \mathbb{Z} \, .
\end{equation}
The full set of mutually consisted labels form a certain lattice $\mathcal{L}_{UV}$ of UV labels \cite{Aharony:2013hda,Gaiotto:2010be}.


In the IR the study of line operators can be understood as a BPS spectral problem. This information is encoded in the quantum line operator
\be \label{QuantumLine}
\mathscr{L}_\wp = \sum_\gamma \ \underline{\overline{\Omega}} (L , u , \gamma ; q) \ \mathsf{X}_\gamma \, ,
\ee
where the refined BPS degeneracies (Protected Spin Characters) are defined as traces over the BPS Hilbert space of states bounded to the line defect
\be \label{PSC}
\underline{\overline{\Omega}} (L ,\gamma , u , \zeta ; q) :=  \Tr_{\cH_{L, \gamma , u , \zeta}} q^{2 J_3} (-q)^{2 I_3} \, .
\ee
Consequently we have the two specializations of \eqref{QuantumLine}
\begin{align}
\langle L_{\zeta , \alpha} \rangle_{q=+1 , u}  &= \sum_{\gamma \in \Gamma_L} \underline{\overline{\Omega}} ( u , L , \gamma ; q = +1) \ Y_{\gamma} \ ,  \label{vevLY}
\\
\langle L_{\zeta , \alpha} \rangle_{q=-1 , u} &= \sum_{\gamma \in \Gamma_L} \underline{\overline{\Omega}} ( u , L , \gamma ; q = -1) \ X_{\gamma} \, .
 \label{vevLX}
\end{align}
We will often omit the $q,u$ labels and let the use of the coordinates $X$ or $Y$ identify the specialization we are talking about. The set of consistent line defects of a given theory forms an algebra, where the product and multiplication are defined via the insertion of the defects in the path integral.

As in the case of BPS quivers, also the line defect spectral problem can be often reformulated in an algebraic language. In the IR a line defect is characterized by an effective quantum mechanical model which describes the low energy dynamics of a cloud of BPS particles bound to an infinitely massive dyonic particle, called \textit{the core} \cite{Cordova:2013bza,Gaiotto:2010be}. This low energy quantum mechanics is associated to the representation theory of a framed BPS quiver $Q[f]$. The latter is obtained by adding to the BPS quiver which describes the ordinary BPS spectrum an extra node $f$ which corresponds to the core charge $e_f$ of the defect, connected to the unframed quiver using the symplectic pairing. The central charge function is extended to $Q[f]$ by linearity. If the addition of the framing node produces new closed cycles, the superpotential of $Q$ can be modified by the addiction of new terms $\cW_L$.

Framed BPS states are now associated to framed representations. There are two natural choices of stability conditions, which select cyclic or co-cyclic representations. In both cases the problem has a combinatorial solution for a large class of quivers. One of the main results of \cite{Cirafici:2019otj} is the identification of the protected spin character \eqref{PSC} with refined Donaldson-Thomas invariants of the moduli space of framed quiver representations

Moving along the Coulomb branch $\mathcal{B}$ it will generically happen that the core charge will change and the framed BPS quiver mutate accordingly. This process can occur at anti-walls, loci where $Z(\gamma)/\xi \in - \ii \mathbb{R}_+$, with $\gamma$ the charge of an ordinary BPS state. These quiver mutations correspond to physical processes, in which the core charge $e_f$ changes due to the fusion/fission with a BPS particle $\gamma$ at the corresponding anti-wall \cite{Cordova:2013bza,Gaiotto:2010be}. Note that anti-walls are in one to one correspondence with the BPS walls.

Line operators are order parameters and since no phase transition occurs in the Coulomb branch the vevs $\langle L_{\zeta,\alpha}\rangle_{q=\pm1}$ are continuous $J_\zeta$-holomorphic functions. Concretely this means that the framed BPS degeneracies undergo wall-crossings in order to compensate the discontinuities in the $X_\gamma$ or $Y_\gamma$ functions, such that the vev remains invariant. 

In particular we can imagine rotating the phase $\zeta$ along the whole upper half plane, crossing all the BPS walls and all the anti-walls. Note that these two operations are physically distinct, but since we are crossing \textit{all} the walls they are operationally the same. Therefore the sequence of  mutations of the framed quiver is precisely induced by the sequence of mutations of the unframed BPS quiver. In the following, since we will only be concerned with sequence of mutations which cover all the BPS spectrum (or in case there is an R-symmetry finer sequences which however contain the same information) we will loosely talk about quiver mutations and cluster transformations without distinguishing the case of framed vs. unframed quivers. A more detailed discussion is in \cite{Cirafici:2017iju}.

\section{Dynamical systems from theories of class $S$} \label{DynSys}

In this Section we introduce certain discrete dynamical systems defined via the composition of cluster transformations. Such dynamical systems act on the space of line operators. They have the form of iterated rational maps which take the vev of a line defect into the vev of another, in general distinct, line defect. The main idea is that we can use this iterated maps to produce new line defects and therefore explore the space of line defects\footnote{Some readers might be more familiar with the analog statement in Morse theory, where by studying a certain continuous dynamical system associated with a Morse function one can hope to gain topological information about the ambient space.}.

%

%

\subsection{Framed quivers and cluster transformations} \label{FQCT}

The main idea behind this paper, as well as \cite{Cirafici:2013bha,Cirafici:2017iju} is this: sequences of quiver mutations which transform a BPS quiver back to itself, in general will not do so when the same quiver is \textit{framed}. Therefore the sequences of quiver mutations $\mathbf{m}^+$ or $\mathbf{r}^+$ act non-trivially on framed quivers. This induces an action of cluster transformations on the set of line operators. This action follows from the fact while mutations change the form of the quiver, a line operator is actually wall-crossing \textit{invariant}. This is enough to constrain the form of other line operators corresponding to mutations of the quiver \cite{Cirafici:2013bha,Cirafici:2017iju}, as we will review momentarily.

Having fixed a positive basis $\{ e_i \}_{i=1}^n$ at a point of $\cB$, any stable framed BPS state has charge of the form $\gamma = e_f +  \sum_{i \in Q_0} \,  d_i \, e_i$, where $e_f$ is the core charge and $d_i \in \mathbb{Z}_{\ge 0}$. We identify the BPS degeneracies with (eventually refined) Donaldson-Thomas invariants of framed quivers, assuming cyclic stability conditions, as in \cite{Chuang:2013wt,Cirafici:2017wlw,Cirafici:2019otj}. Therefore for a line defect labelled by $\alpha$ in the UV, we can write
\begin{equation}
\langle L_{\zeta , \alpha} \rangle_{q=+1} = \sum_{\mathbf{d} \in \mathbb{Z}_{\ge 0}^{|Q_0|}} \ \DT (Q[f] , \cW ) \ Y_{e_f + d_i \, e_i} (\zeta) 
\, .
\end{equation}
We stress that from this formula it is clear that the degeneracies only depend on the framed quiver and its superpotential (having fixed stability conditions). 

Consider now a different line defect $L_\beta$, described by the new framed quiver $Q[f']$ with superpotential $\cW'$, at the same point in the moduli space. As before
\begin{equation} \label{Lbeta}
\langle L_{\zeta , \beta} \rangle_{q=+1} = \sum_{\mathbf{d} \in \mathbb{Z}_{\ge 0}^{|Q_0|}} \ \DT (Q[f'] , \cW' ) \ Y_{e_{f'} + d_i \, e_i} (\zeta) 
\, .
\end{equation}

Assume now that the sequence of mutations $\mathbf{m}^+$ which scans over all the vanilla BPS spectrum, acts on the framed quiver by taking $(Q[f],\cW)$ in $( Q[f'] , \cW')$. As we cross the full set of BPS walls (and therefore the anti-walls), the framed BPS quiver undergoes quiver mutations and the cluster coordinates jump according to \eqref{clusterR}. After scanning all of the vanilla BPS spectrum, the BPS quiver will be back to itself, eventually up to a permutation of the nodes. To simplify the formulae let us assume for the moment that this permutation is the identity. The corresponding transformation on the $Y_\gamma$ coordinates is the action of the rational transformation $R_{\mathbf{m}^+}$ associated with the half-monodromy. By the wall-crossing invariance of $L_\alpha$ the vevs have no discontinuities and we have
\begin{align} \label{WCcond}
\langle L_{\zeta , \alpha} \rangle_{q=+1} & = \sum_{\mathbf{d} \in \mathbb{Z}_{\ge 0}^{|Q_0|}} \ \DT (Q[f] , \cW ) \ Y_{e_f + d_i \, e_i} 
\\ \label{ECcond}
&  = \sum_{\mathbf{d} \in \mathbb{Z}_{\ge 0}^{|Q_0|}} \ \DT (Q[f'] , \cW' ) \ R_{\mathbf{m}^+} \, [Y_{e_f + d_i \, e_i} ] = \langle L_{-\zeta , \alpha} \rangle_{q=+1} 
\end{align}
Note that for the equality to hold, we have to express the mutated cluster variables in terms of the original seed; otherwise we simply have two different functions of two different seeds $\{ Y_i \}$ and $\{ Y_i' \}$.

The crucial observation is that the framed BPS degeneracies, since they are DT invariants of the framed quiver, \textit{do not depend on the cluster coordinates}. In particular $\DT (Q[f'], \cW')$ are associated with the quiver $Q[f']$ \textit{regardless} whether we see this quiver as obtained from wall crossings from $Q[f]$ or as representing a physically distinct operator.

We can regard \eqref{WCcond} as a condition which determines $\DT (Q[f'],\cW')$ whenever the degeneracies $\DT (Q[f],\cW)$ and the rational transformation $R_\mathbf{m^+}$ are known. Having found the degeneracies  $\DT (Q[f'],\cW')$ we simply plug then in \eqref{Lbeta} and obtain the vev $\langle L_{\zeta , \beta} \rangle_{q=+1} $ of a physically distinct line operator at the same point of the moduli space where we knew $\langle L_{\zeta , \alpha} \rangle_{q=+1} $.

Equivalently this condition can be recast as \cite{Cirafici:2013bha}
\begin{align} \label{gensolL}
\langle L_{\zeta , \beta} \rangle_{q=+1} &=  \sum_{\mathbf{d} \in \mathbb{Z}_{\ge 0}^{|Q_0|}} \ \DT (Q[f'],\cW') \ Y_{e_{f'} + d_i \, e_i} \, , \\
& = \sum_{\mathbf{d} \in \mathbb{Z}_{\ge 0}^{|Q_0|}} \ \DT (Q[f],\cW) \ R^{-1}_{\mathbf{m}^+} \, [ Y_{e_{f'} + d_i \, e_i} ] \, ,
\end{align}
which is more practical for computations.

Let us rephrase the argument in a more compact way. Let us denote by $F \left[Q [f]; \{ Y_1,\dots,Y_n \} \right]$ the function expressing the vev of a line operator associated with the quiver $Q[f]$ in terms of the seed $\{ Y_1 , \dots Y_n \}$. Then the invariance under framed wall-crossing of the line operators vev's is described by the identity
\begin{equation}
F\left[Q [f]; \{Y_1,\dots,Y_n \} \right] = F \left[ \mathbf{m}^+ Q [f]; R_{\mathbf{m}^+} \{ Y_1,...,Y_n \} \right] \ .
\end{equation}
By relabelling the seed this is mathematically equivalent to
\begin{equation} \label{gensolF}
F \left[ \mathbf{m}^+ Q[f] ; \{ Y_1,\dots,Y_n\} \right] = F \left[Q[f] ; R_{\mathbf{m}^+}^{-1} \{ Y_1,...,Y_n \} \right] \, ,
\end{equation}
which however physically expresses the framed degeneracies of a line operator described by $Q[f']=\mathbf{m}^+ Q[f]$ in terms of those of a different line operator associated to $Q[f]$, at the same point in the moduli space. The line defect vev $\langle L_{\zeta , \beta} \rangle_{q=+1} $ is obtained from $\langle L_{\zeta , \alpha} \rangle_{q=+1} $ by applying the opposite sequence of cluster transformations $R^{-1}_{\mathbf{m}+}$ on the initial seed $\{ Y_1 , \dots , Y_n \}$. This arguments generalize to the case where the quiver $Q$ goes back to itself after the sequence of mutations $\mathbf{m}$ only when composed with a non-trivial permutation $\sigma$. In this case the same arguments hold word by word and \eqref{gensolF} is replaced by
\begin{equation} \label{gensolFfractional}
F \left[\mathbf{m}^+ Q[f] ; \{ Y_1,\dots,Y_n \} \right] = F \left[ Q[f] ; (\sigma \circ R_{\mathbf{m}^+})^{-1} \{ Y_1,\dots,Y_n \} \right] \, .
\end{equation}
These arguments extend also to the case where a chamber has a $\mathbb{Z}_{2 s}$ discrete symmetry induced by a residual $R$-symmetry which acts on the central charges as $Z \longrightarrow \e^{\ii \pi /s} Z$ and the sequence $\mathbf{r}^+$ is a fractional monodromy.

The same arguments can be rephrased in terms of the $\tilde{R}$ operations, which indeed coincide with the $R$ transformations for the opposite quiver. Furthermore in many cases, for example when the elementary rational transformations in a fractional monodromy commute (as is the case when the sequence of mutations involves only non adjacent nodes) the elementary transformations $\tilde{R}_j^{i_k}$ and $R_j^{i_k}$ are inverse to each other and one can reformulate \eqref{gensolFfractional} using only  $\tilde{R}_j^{i_k}$, as in \cite{Cirafici:2013bha,Cirafici:2017iju}.


\subsection{Discrete dynamical systems from cluster transformations}

The action of quiver mutations and cluster transformations on the set of line operators defines naturally a discrete dynamical system, as the iteration of the above defined rational transformations $R$.

\paragraph{Discrete dynamical systems.}
%
%
%
%
%
%

Consider a set $S$ and let $f : S \longrightarrow S$ a map. In general one can think of $f$ as a process which takes an initial state $x_0 \in S$ in a new state $x_1 = f (x_0)$ after one iteration of the map. One can iterate the map an arbitrary number of times, and we denote by $f^n := f \circ \cdots \circ f$ its $n$-th iteration. We say that $x_n$ is the $n$-th image of the initial condition $x_o$ under $f$ if $x_n = f^n (x_0)$. We define the \textit{forward orbit} of $f$ as $\cO^+ (x_0) := \{ x_n \}_{n \in \mathbb{N}}$. If the map is invertible, we can similarly iterate its inverse $f^{-1}$ and define $x_n := f^{-n} (x_0)$. Then we define the \textit{full orbit} of $x_0$ as the set $\cO (x_0) = \{ x_n \}_{n \in \zed}$.

The simplest example of an orbit for $f$ is a \textit{fixed point} $x_0$, for which $f(x_0) = x_0$. If $f$ is an invertible map, we have that $\cO (x_0) = \{ x_0 \}$. If $f^k (x_0) = x_0$ we say that $x_0$ is a \textit{periodic point} of period $k \ge 1$. In particular a fixed point is also a periodic point of every period. If $x_0$ is a periodic point of period $k$, it is also a periodic point of period $n \, k$ for any integer $n \ge 1$. Therefore it is useful to introduce the concept of \textit{minimal period} $k\ge1$, the minimal value so that $f^k (x_0) = x_0$.

\paragraph{The action on line operators}

Finally we can put everything together. Denote by $\mathsf{Line}$ the abstract set of line operators (this is actually an algebra, but we will not really need the algebraic structure). Similarly we denote by $\mathsf{Line}_q$ the set of quantum line operators, whose elements are in one to one correspondence with the elements of $\mathsf{Line}$. All such operators can be understood as sums over the charge lattice of monomials in the (quantum) cluster variables, whose coefficients are the (refined) framed BPS states degeneracies, assuming no BPS particle transforms in a non trivial representation of the $\mathfrak{su}(2)_R$ R-symmetry group \cite{Gaiotto:2010be}. Having fixed a basis $\{ e_i \}$, we regard $\mathsf{Line}$ as a subset of the ring of Laurent polynomials $R [\{ Y_i \} , \{ Y_i^{-1} \}]$. Analog considerations hold for quantum line operators, where now we consider a $q$-deformation of $R [\{ Y_i \} , \{ Y_i^{-1} \}]$.

Consider first the classical case, in terms of the twisted variables $\{ Y_i \}$. As we have discussed the rational transformations corresponding to the quantum monodromies act on these variables. However we have just shown that this action induces an action on line operators. Therefore we have implicitly constructed a rational map, which acts on $\mathsf{Line}$ by repeated iterations, that is a discrete dynamical system. We write this implicitly as
\be
\mathsf{R} \, : \, \mathsf{Line} \longrightarrow \mathsf{Line} \, ,
\ee
where now $\mathsf{R}$ is the map induced by the rational transformations \eqref{RonCluster} and \eqref{RtonCluster} according to the rules \eqref{gensolFfractional}. This map has to be understood as taking as input an IR line operator, written as an expansion in cluster variables at a point $u \in \cB$ where this expansion is valid, and giving a different line operator at the same point in the Coulomb branch and in the same cluster coordinates\footnote{Note that this dynamical system is different from the one obtained by considering the iterated cluster transformations on the cluster variables, studied for example in \cite{Neitzke:2017cxz}. Our $\mathsf{R}$ acts on the set of line defects.
}. 

Note that the technical operations to define this dynamical system are not different from the rational operations we have discussed in Section \ref{classS}. However the conceptual perspective is different, and richer. We can now use all the tools at our disposal in the theory of dynamical systems to study the topology and geometry of the set $\mathsf{Line}$. The first step in doing so is to characterise orbits and fixed points of $\mathsf{R}$.

Similarly we can define a quantum dynamical system
\be
\mathsf{R}_q \, : \, \mathsf{Line}_q \longrightarrow \mathsf{Line}_q \, ,
\ee
which now acts by conjugation by the quantum monodromy. This object is significantly more difficult to study since rational transformations now depend on noncommutative variables. To the best of our knowledge there is no discussion in the literature on dynamical systems of an iterated map based on coordinates with non-trivial noncommutative relations. We leave this problem for the future.

\paragraph{Remark.} Note that the above definition of the dynamical system implicitly contains the information about a BPS chamber on the Coulomb branch. Specifying a chamber is equivalent to the specification of a stability condition, which can be thought of as an ordering of the phases of the central charges of all the BPS states corresponding to the nodes of the quiver, including the framed node. In this paper we always take the phase of the core BPS state in the defect to be much bigger than that of all the other particles, while keeping the mass of the defect much bigger than all the other masses. Mathematically this correspond to selecting cyclic framed quiver representations, as discussed in Section \ref{classS}. What remain to be fixed is the ordering of the arguments of the central charge computed on the BPS particles corresponding to the nodes of the unframed quiver. This is implicitly chosen when defining the dynamical system via a sequence of quiver mutations which take the framed quiver to another framed quiver while taking the unframed quiver back to itself. The sequence of mutations corresponds to an ordering of such phases and therefore to a BPS chamber in the Coulomb branch. Note that while the full rational transformation is independent on the position on the Coulomb branch, the specific sequence of mutation provides the information about which framed quivers, and therefore which core charges of line operators, one can obtain.

\paragraph{An example.} We will begin with a simple example in the framework of $A_2$ Argyres-Douglas theory. The framed BPS spectrum is well known \cite{Gaiotto:2010be} and simple enough to illustrate our points without too much algebra. The BPS quiver is
\begin{equation}
\xymatrix@C=8mm{ 
 \circ  \ar[r] &  \bullet    
 } \, .
\end{equation}
We choose a point in the Coulomb branch with a $\mathbb{Z}_4$ symmetric spectrum, consisting of particles with charges $e_\bullet$ and $e_\circ$ and their anti-particles. This point lies within a certain chamber, where our results hold. In this case the $\mathbf{r}^+$ sequence is given by $\bullet$, with permutation $\sigma = ( \bullet , \circ )$. Equivalently the half-monodromy is constructed out of the sequence $\{ \bullet , \circ \}$. We have the rational transformations associated with $\mathbf{r}^+$ :
\be \label{RtransfA2}
R \equiv
\begin{cases}
&Y_\bullet\to 1/Y_\bullet \\
&Y_\circ\to  (1 + Y_\bullet) \, Y_\circ
\end{cases}
\ , \qquad
\tilde{R}\equiv
\begin{cases}
&Y_\bullet\to 1/Y_\bullet \\
&Y_\circ\to Y_\bullet \, Y_\circ / (1 + Y_\bullet)
\end{cases} \ .
\ee
Note that the composition $\tilde{R} \circ R = \mathrm{id}$, and therefore $R^{-1} = \tilde{R}$. 

In this example we can give a graphical representation of our discrete dynamical system in terms of the framed BPS quiver. Recall that a defining property of our dynamical system is that it sends the BPS quiver back to itself, eventually by relabelling the nodes, but a framed BPS quiver representing a line operator into a different framed BPS quiver.

In this case the sequence of mutations is simply $\sigma \circ \mu_\bullet$ and by applying it repeatedly we find the following orbit of framed BPS quivers
\begin{eqnarray}
\begin{matrix}
\xymatrix@C=8mm{   f_{e_\bullet}  & \\
 \circ \ar[u]  \ar[r] &  \bullet    
 }
\end{matrix}
\begin{array}{c} \sigma \circ \mu_\bullet \\ \Longrightarrow \end{array}
\begin{matrix}
\xymatrix@C=8mm{  &  f_{-e_\circ}  \\
 \circ  \ar[r] &  \bullet   \ar[u] 
 }
\end{matrix}
\begin{array}{c} \sigma \circ \mu_\bullet \\ \Longrightarrow \end{array}
\begin{matrix}
\xymatrix@C=8mm{    f_{-e_\bullet - e_\circ}  \ar[d] & \\
 \circ  \ar[r] &  \bullet    \ar[ul]
 }
\end{matrix}
%
\begin{array}{c} \sigma \circ \mu_\bullet \\ \Longrightarrow \end{array}
\begin{matrix}
\xymatrix@C=8mm{   f_{-e_\bullet}  \ar[d] & \\
 \circ  \ar[r] &  \bullet    
 }
\end{matrix}
\begin{array}{c} \sigma \circ \mu_\bullet \\ \Longrightarrow \end{array}
\begin{matrix}
\xymatrix@C=8mm{  &  f_{e_\circ} \ar[d] \\
 \circ  \ar[r] &  \bullet    
 }
\end{matrix} 
%
\end{eqnarray}
The orbit has period five, $ (\sigma \circ \mu_\bullet)^5 Q[f_{e_\bullet}] = Q[f_{e_\bullet}] $. The framed quivers listed above correspond to the line defects \cite{Gaiotto:2010be,Cecotti:2010fi}
\begin{eqnarray}
F[Q  [  f_{e_\bullet} ] ; \{Y_\circ , Y_\bullet \}]  &=& Y_\bullet \, , \\
F[Q  [f_{-e_\circ} ] ; \{Y_\circ , Y_\bullet \}]  &=&\frac{1}{Y_\circ}  \, , \\
F[Q  [f_{-e_\bullet - e_\circ}] ; \{Y_\circ , Y_\bullet \}]  &=& \frac{1}{Y_\bullet} + \frac{1}{Y_\bullet Y_\circ} \, , \\
F[Q  [f_{-e_\bullet} ] ; \{Y_\circ , Y_\bullet \}]  &=& \frac{1}{Y_\bullet} + \frac{Y_\circ}{Y_\bullet} + Y_\circ \, , \\
F[Q  [f_{e_\circ}] ; \{Y_\circ , Y_\bullet \}]  &=& Y_\circ + Y_\bullet Y_\circ \, .
\end{eqnarray}
We have temporarily adopted the notation of \eqref{gensolFfractional} to make the argument clearer. Now, according to \eqref{gensolFfractional}, acting repeatedly with $(\sigma  \circ R)^{-1} $ on all the cluster variables in the above sums of monomials defines the action of our dynamical system map $\mathsf{R}$ on the set of line operators. Since the inverse of $R$ is $\tilde{R}$ in \eqref{RtransfA2}, we find
\begin{equation}
\begin{matrix}
\xymatrix@C=7mm{ \cdots \ar[r]^{\! \! \! \mathsf{R}} & \, Y_\bullet \,
 \ar[r]^{\! \! \! \mathsf{R}} & \,  1/ Y_\circ  
 \ar[r]^{\hspace{-0.9cm} \mathsf{R}} & \, 1/Y_\bullet + 1/ Y_\bullet Y_\circ 
 \ar[r]^{\hspace{-0.5cm} \mathsf{R}} & \, 1/Y_\bullet + Y_\circ / Y_\bullet + Y_\circ
 \ar[r]^{\hspace{0.6cm} \mathsf{R}} & \, Y_\circ + Y_\bullet Y_\circ 
 \ar[r]^{\hspace{0.6cm} \mathsf{R}} & \cdots
 }
\end{matrix}  \ .
\end{equation}
This is a concrete example of a finite orbit for the dynamical system $\mathsf{R}$.

\paragraph{Dynamical systems from theories of class $S$.}

In the rest of the paper will discuss a few physical applications of the iterated map $\mathsf{R}$, with several examples. In abstract terms, we can draw the following consequences
\begin{itemize}
\item Line operators in theories of class $S$ are organized in orbits under the action of $\mathsf{R}$. If a vev of a representative of the orbit is known, it is possible, even if technically challenging, to compute \textit{all} the other operators in the same orbit by iterating the map $\mathsf{R}$.
\item Such orbits can be finite, that is periodic orbits containing a finite set of elements, or infinite. Finite orbit occur every time the quantum monodromy has finite order, that is we can find an integer $\ell$ such that $\mathscr{M}_q^\ell = 1$. Indeed since by definition the dynamical system is determined by its iteration, $\ell$ determines the period (and analog statements if there are fractional monodromies). Finite order orbits appear for example in superconformal theories of Argyres-Douglas type \cite{Cirafici:2013bha}, as we have seen above.
\item Let us take a Lie algebra $\mathfrak{g}$. Then we can construct several quantum field theories from $\mathfrak{g}$ which are distinguished by the set of UV admissible labels for the line operators \cite{Aharony:2013hda}. The IR counterpart of this statement is that we can distinguish all the possibilities by studying the orbits of $\mathsf{R}$. We do this in an example in Section \ref{reading}.
\item Fixed points of $\mathsf{R}$ are somewhat singled out. For example in the case of pure asymptotically free gauge theories they correspond to purely electric Wilson lines; whenever the adjacency matrix is invertible they can be identified with the conserved charges of an associated Q-system \cite{Cirafici:2017iju} (see \cite{Cecotti:2014zga} for a review). In a sense our results generalize \cite{Cirafici:2017iju} to the case where the adjacency matrix is not invertible and no associated Q-system is available. The condition that a line operator is a fixed point of $\mathsf{R}$ has the form of a difference equation in the discrete dynamical system. Sometimes this equation can be solved explicitly to compute the line operator vev. In Section \ref{sDAHA} we show explicitly some fixed points in the case of  superconformal SU(2) Yang-Mills with four flavours. In the quantum version we prove by direct computation that they are generators of the spherical subalgebra of a Double Affine Hecke Algebra.
\item On the other hand sometimes we know indirectly that some operators are fixed points and we can use this fact to compute explicitly $\mathsf{R}$ as a rational transformation. We do so in the case of SU(2) $\cN=2^*$, where the spectrum generator is unknown in closed form, in Section \ref{fixedpoints}.
\end{itemize}

%

\section{Global structures from orbits} \label{reading}

In this Section we will explain how our dynamical system can be used to ``read between the lines'', in the sense of \cite{Aharony:2013hda} in the IR limit of gauge theories with a lagrangian description. The dynamical system is defined via the local theory but acting on mutually non compatible line operators produce distinct orbits. We will show this with a detailed example.

\subsection{Line defects, quantum monodromies and R-symmetries}

The existence of a fractional monodromy is a consequence of an unbroken R-symmetry which acts on the central charge by multiplication by a phase. In this case the spectrum has a $\zed_{2s}$ discrete symmetry, induced by the action of a discrete R-symmetry on the central charge operator $Z \longrightarrow \e^{i \pi /s} Z$. Such an R-symmetry is related to the periodicity of the $\theta$-angle. Recall that if we consider a theory in an instanton background of instanton number $k$, then by the index theorem the net number of fermionic zero modes for a fermion in the representation $\mathbf{r}$ is expressed in terms of the quadratic Casimir of the representation as $2 \, C_2 (\mathbf{r}) \, k$. As a consequence under a phase rotation $\psi \longrightarrow \e^{\ii \delta} \psi$ the measure of the path integral changes by a phase $\e^{2 C_2 (\mathbf{r}) k \ii \delta}$. Therefore if we now consider an R-symmetry, it will be an actual quantum symmetry of the path integral if we can remove the extra phase factor with a $\theta$-angle rotation. As a consequence different periodicities of the $\theta$-angle, for example as in an SU(2) vs an SO(3) theory, determine the existence or not of a fractional quantum monodromy. Note that the rational transformation corresponding to the half-monodromy is the same in the SU(2) and SO(3) theories, and in purely mathematical terms can always be decomposed in a $1/4$-monodromy. However this decomposition is unphysical for SO(3) and physical only for SU(2).

More generally if we have a theory based on the gauge group $G$ with Lie algebra $\mathfrak g$, we can always write $G = \tilde{G} / {\mathbf{H}}$ where $\mathbf{H} \subset \mathbf{C}$, with $\tilde{G}$ the universal covering group, and $\mathbf{C}$ its center. If the group $G$ is not simply connected then the instanton number can be fractional in the presence of a line operator. This  fact, as well as the possible presence of other characteristic classes, specify the periodicity of the $\theta$-angle \cite{Aharony:2013hda}. Having established what is the smallest physical fractional monodromy we can use it to define our dynamical system $\mathsf{R}$. 

Now to construct the full set of allowed line operators we start from a chamber with a basis $\{ e_i \}$ of $\Gamma$ and construct core charges in an extension of $\Gamma$, with the condition that they have integral pairing with the element of the basis and between each other. From each of this core charges, assuming we can compute its line operator (for example with localization techniques \cite{Cirafici:2017wlw,Cirafici:2019otj}) we iterate the map $\mathsf{R}$ to generate a full orbit. The relation between $\mathsf{R}$ and wall-crossing (or equivalently the rules of quiver mutations) ensures that each element of the orbit is an allowed line operator, whose core charge obeys the Dirac quantization condition. Therefore we can talk of compatible orbits, if the core charge of any two representatives have integral pairing. In other words what in the UV is the classification of allowed Lie algebra weights $(\lambda_e , \lambda_m)$, in the IR becomes the classification of the allowed orbits of $\mathsf{R}$.

\subsection{The case of $\mathrm{SU(3)}/\mathbb{Z}_3$}

As an example of these considerations we will now consider $\mathrm{SU(3)}/\mathbb{Z}_3$. This case admits three different theories $\left( \mathrm{SU(3)} / \mathbb{Z}_3 \right)_{n}$, $n=0,1,2$ \cite{Aharony:2013hda}. All these theories inherit the set of line defects of the pure SU(3) gauge theory which survive the $\zed_3$ projection. For example their Wilson lines have charges of the form $-m_1 (e_{\bullet_1} + e_{\circ_1}) -m_2 (e_{\bullet_2}+e_{\circ_2})$, with $m_1 , m_2 \in \zed$. We want to study the additional line operators which distinguish these three theories from the point of view of the orbits of the associated dynamical system $\mathsf{R}$.

The pure SU(3) gauge theory has an $\mathbb{Z}_6$ R-symmetry associated with the $\theta \longrightarrow \theta + 2 \pi$ periodicity of the theta angle. This is not anymore the case for $\mathrm{SU(3)} / \zed_3$ theories. Indeed in this case we can have bundles with $\frac{1}{3} \mathbb{Z}$-valued instanton numbers. 
On a generic four manifold $M$, $SU(N)/\mathbb{Z}_N$ bundles are classified by their instanton number $k$ and by their Stiefel-Whitney class $v \in H^2 (M , \mathbb{Z}_N)$ which measures the obstruction to the lift from an $SU(N)$ bundle (and which is $\mathbb{Z}_N$-valued since $\pi_1 (SU(N) / \mathbb{Z}_N) = \mathbb{Z}_N$). The two are however related as $k = \frac{v \smile v}{2 N} - \frac{v \smile v}{2} \, \mathrm{mod} \, 1$. In particular if $M$ is spin, then the cup product $v \smile v$ is even and $k \in \frac1N \mathbb{Z}$.
As a result the periodicity of the theta angle is extended and the shift $\theta \longrightarrow \theta + 2 \pi$ is not anymore a symmetry of the theory, but will map the set of line operators of a theory to the set of line operators of another theory \cite{Aharony:2013hda}.

The SU(3) theory has BPS quiver 
\begin{align}
\xymatrix@C=6mm{
& \bullet_1 \ar[drr] & & \bullet_2 \ar[dll] 
  \\
& \circ_1 \ar@<-0.5ex>[u]  \ar@<0.5ex>[u]  & & \circ_2  \ar@<-0.5ex>[u]  \ar@<0.5ex>[u] 
}
\end{align}
and the BPS spectrum is generated by the sequence of mutations $\mathbf{m}^+= \mu^+_{\bullet_2} \,  \mu^+_{\bullet_1} \,  \mu^+_{\circ_2} \,  \mu^+_{\circ_1} \,  \mu^+_{\bullet_2} \,  \mu^+_{\bullet_1}$. This sequence of mutations implicitly selects a chamber in the moduli space with finite spectrum, given by $\{ \gamma_{\bullet_1} , \gamma_{\bullet_2} , \gamma_{\bullet_2} + \gamma_{\circ_1} , \gamma_{\bullet_1} + \gamma_{\circ_2} , \gamma_{\circ_2} , \gamma_{\circ_1} \}$, plus anti-particles. Our results will hold in this chamber. 

The pure SU(3) theory admits a $1/6$-monodromy, generated by the sequence $\mathbf{r}^+ = \mu^+_{\bullet_2} \, \mu^+_{\bullet_1}$, with permutation $\sigma = \{ (\bullet_1 , \circ_1) , (\bullet_2 , \circ_2) \}$. The existence of such a monodromy is a consequence of an unbroken $\mathbb{Z}_6$ $R$-symmetry, which is however broken in $\mathrm{SU(3)}/\mathbb{Z}_3$. To such a monodromy we can associate a rational transformation $R$ as in Section \ref{DynSys} and then take its inverse, taking into account the permutation:
\begin{equation} \label{RtopSU3}
\begin{cases}
  &Y_{\bullet_1}  \to 1/ Y_{\circ_1}
     \\
   & Y_{\bullet_2} \to  1/Y_{\circ_2}
         \\
   & Y_{\circ_1} \to  \frac{Y_{\circ_1}^2 (Y_{\circ_2}+1) Y_{\bullet_1}}{(Y_{\circ_1}+1)^2}
       \\
   & Y_{\circ_2} \to  \frac{(Y_{\circ_1}+1) Y_{\circ_2}^2 Y_{\bullet_2}}{(Y_{\circ_2}+1)^2}
      \end{cases}   \, .
\end{equation}
This is the physical transformation for the $\mathrm{SU} (3)$ gauge theory. However in the case of $\mathrm{SU(3)}/\mathbb{Z}_3$ the R-symmetry is broken and this transformation becomes unphysical: the physical half-monodromy transformation is given by iterating three times the map \eqref{RtopSU3}. This defines our dynamical system $\mathsf{R}$. We will now show explicitly that this dynamical system generates mutation orbits which are not mutually consistent when starting from non mutually consistent defects; and can therefore be used to detect the three different quantum field theories labelled $\left( \mathrm{SU(3)} / \mathbb{Z}_3 \right)_{n}$. At the end we will comment on the action of the unphysical 1/6 monodromy on this set of line defects.

\underline{$\left( SU(3) / \mathbb{Z}_3 \right)_0$}

We take this theory to be the one with line defects corresponding to the core charges $\frac13 e_{\bullet_1} + \frac23 e_{\bullet_2}$ and $\frac23 e_{\bullet_1} + \frac13 e_{\bullet_2}$. Their vevs can be computed immediately by localization since there are no cyclic modules and the only non-trivial framed BPS state is given by the core charge. This charges are mutually consistent and are consistent with the basis of charges associated with the nodes of the BPS quiver. They in general violate the Dirac quantization condition with an arbitrary Wilson line of SU(3). As a result the set of allowed Wilson lines in this theory is reduced\footnote{The center of SU(N) is $\mathbb{Z}_N$. Since $\mathbb{Z}_N$ has only a finite number of representations, the representations of SU(N) are divided into $N-1$ classes which correspond to the representations of $\mathbb{Z}_N$. These classes can be labelled by their $N$-ality, given by the number of boxes in the corresponding Young tableau $\mathrm{mod} \, N$. Taking the quotient by  $\mathbb{Z}_N$ means that we keep only those representations of SU(N) which have $0 \, \mathrm{mod} \, N$ $N$-ality. In the case of $SU(3) / \mathbb{Z}_3$ these are the $\mathbf{8}$, the $\mathbf{10}$ and so on.} and corresponds precisely to the representations of $SU(3)$ which are invariant under the center $\mathbb{Z}_3$. On the other hand the set of dyonic line defects of the pure $SU(3)$ theory constructed in \cite{Cirafici:2017iju} is still present. The charges of such operators have all integral coefficients in the BPS quiver basis $\{ e_i \}_{i \in Q_0}$, and therefore the new core charges have integral pairing with this set. 

Out of these charges we can construct whole families of line operators by iterating the map $\mathsf{R}$. For example the line defect with core charge $\frac13 e_{\bullet_1} + \frac23 e_{\bullet_2}$ has vev $Y_{\frac13 e_{\bullet_1}} \, Y_{\frac23 e_{\bullet_2}}$. By iterating the discrete map $\mathsf{R}$ we can construct the forward orbit $\cO^+ \left( \langle L_{\frac13 e_{\bullet_1} + \frac23 e_{\bullet_2}} \rangle \right) $ and the backward orbit $\cO^- \left( \langle L_{\frac13 e_{\bullet_1} + \frac23 e_{\bullet_2}} \rangle \right) $. We show a sample of the full orbit : 
\begin{eqnarray}
& \ \vdots \ & \cr
\langle L_{- \frac23 e_{\bullet_1} - \frac13 e_{\bullet_2}} \rangle &=& \frac{1}{Y_{e_{\bullet_1}}^{2/3} Y_{e_{\bullet_2}}^{1/3}}\Big(1+Y_{e_{\circ_1}}+2 Y_{e_{\bullet_1}} Y_{e_{\circ_1}}+Y_{e_{\bullet_1}}^2 Y_{e_{\circ_1}}+2 Y_{e_{\bullet_1}} Y_{e_{\circ_1}} Y_{e_{\circ_2}}   \cr && \ \ \  +2 Y_{e_{\bullet_1}}^2 Y_{e_{\circ_1}} Y_{e_{\circ_2}}+2 Y_{e_{\bullet_1}} Y_{e_{\bullet_2}} Y_{e_{\circ_1}} Y_{e_{\circ_2}}  +2 Y_{e_{\bullet_1}}^2 Y_{e_{\bullet_2}} Y_{e_{\circ_1}} Y_{e_{\circ_2}}  \cr && \ \ \  +Y_{e_{\bullet_1}}^2 Y_{e_{\circ_1}} Y_{e_{\circ_2}}^2 +3 Y_{e_{\bullet_1}}^2 Y_{e_{\bullet_2}} Y_{e_{\circ_1}} Y_{e_{\circ_2}}^2   \cr && \ \ \  +3 Y_{e_{\bullet_1}}^2 Y_{e_{\bullet_2}}^2 Y_{e_{\circ_1}} Y_{e_{\circ_2}}^2+Y_{e_{\bullet_1}}^2 Y_{e_{\bullet_2}}^3 Y_{e_{\circ_1}} Y_{e_{\circ_2}}^2\Big) \, , 
\\
\langle L_{\frac13 e_{\bullet_1} + \frac23 e_{\bullet_2}} \rangle &=& Y_{e_{\bullet_1}}^{1/3} Y_{e_{\bullet_2}}^{2/3}  \, , \\
\langle L_{- \frac23 e_{\bullet_1} - \frac43 e_{\bullet_2} - e_{\circ_1} -2 e_{\circ_2}} \rangle &=& \frac{1+2 Y_{e_{\circ_2}}+Y_{e_{\circ_2}}^2+Y_{e_{\bullet_2}} Y_{e_{\circ_2}}^2+Y_{e_{\bullet_2}} Y_{e_{\circ_1}} Y_{e_{\circ_2}}^2}{Y_{e_{\bullet_1}}^{2/3} Y_{e_{\bullet_2}}^{4/3} Y_{e_{\circ_1}} Y_{e_{\circ_2}}^2} \, , \\
& \ \vdots \ &
\end{eqnarray}
Overall elements of this family have core charges
\begin{equation}
\begin{cases}
& \left(-\frac23 (3n+1) e_{\bullet_1} - \frac{3n+1}{3} e_{\bullet_2} - 2n e_{\circ_1} - n e_{\circ}\right)_{n \ge 0} \\ 
& \left( - \frac{3n-1}{3} e_{\bullet_1} - \frac23 (3n-1) e_{\bullet_2} - n e_{\circ_1} - 2 n e_{\circ_2} \right)_{n \ge 0}
\end{cases} \, .
\end{equation}
Similarly from the defects with core charge $\frac23 e_{\bullet_1} + \frac13 e_{\bullet_2}$ we generate the families of defects with core charges
\begin{equation}
\begin{cases}
& \left( - \frac{3 n+1}{3} e_{\bullet_1} - \frac23 (3n+1) e_{\bullet_2} - n e_{\circ_1} - 2 n e_{\circ_2}\right)_{n \ge 0} \\
& \left( -\frac23 (3n-1) e_{\bullet_1} - \frac{3n-1}{3} e_{\bullet_2} - 2n e_{\circ_1} - n e_{\circ_2} \right)_{n \ge 0}
\end{cases} \, .
\end{equation}
Of course this is not the full set of line defects, but just a few very simple orbits. The explicit form of the vevs for these defects, and other elements of the orbits, may contain hundred of terms; therefore we don't write down here and refer the reader to the accompanying \textsc{mathematica} file \cite{mathe}.

\underline{$\left( SU(3) / \mathbb{Z}_3 \right)_1$}

Now we consider the theory which contains line defects with core charge $-\frac13 e_{\circ_1} - \frac23 e_{\circ_2}$ and $-\frac23 e_{\circ_1} - \frac13 e_{\circ_2}$. Again these charges are a consistent set within themselves and with the basis of charges of the BPS quiver. The set of dyonic line defects of the original $SU(3)$ theory is still present, and the set of Wilson lines is reduced to those in representations of $SU(3)$ which are invariant under $\mathbb{Z}_3$. However now if we begin from the core charge $-\frac13 e_{\circ_1} - \frac23 e_{\circ_2}$, by iterating the rational transformation corresponding to the half-monodromy, we get the line defects
\begin{eqnarray}
& \ \vdots \ & \cr
\langle L_{\frac23 e_{\circ_1}+ \frac13 e_{\circ_2}} \rangle &=& Y_{e_{\circ_1}}^{2/3} Y_{e_{\circ_2}}^{1/3} \left(1+Y_{e_{\bullet_1}}+Y_{e_{\bullet_1}} Y_{e_{\circ_2}}+2 Y_{e_{\bullet_1}} Y_{e_{\bullet_2}} Y_{e_{\circ_2}}+Y_{e_{\bullet_1}} Y_{e_{\bullet_2}}^2 Y_{e_{\circ_2}}\right) \, , \cr
\langle L_{\frac13 e_{\circ_1} + \frac23 e_{\circ_2}} \rangle &=& \frac{1}{Y_{e_{\circ_1}}^{1/3} Y_{e_{\circ_2}}^{2/3}} \, , \cr
\langle L_{-e_{\bullet_1}-2 e_{\bullet_2}-\frac43 e_{\circ_1}-\frac83 e_{\circ_2}} \rangle &=& \frac{1}{Y_{e_{\bullet_1}} Y_{e_{\bullet_2}}^2 Y_{e_{\circ_1}}^{4/3} Y_{e_{\circ_2}}^{8/3}}\Big(1+3 Y_{e_{\circ_2}}+3 Y_{e_{\circ_2}}^2+2 Y_{e_{\bullet_2}} Y_{e_{\circ_2}}^2+2 Y_{e_{\bullet_2}} Y_{e_{\circ_1}} Y_{e_{\circ_2}}^2
\cr & & \ \  +Y_{e_{\circ_2}}^3+2 Y_{e_{\bullet_2}} Y_{e_{\circ_2}}^3  +Y_{e_{\bullet_2}}^2 Y_{e_{\circ_2}}^3 +2 Y_{e_{\bullet_2}} Y_{e_{\circ_1}} Y_{e_{\circ_2}}^3
+2 Y_{e_{\bullet_2}}^2 Y_{e_{\circ_1}} Y_{e_{\circ_2}}^3 \cr  & &  \ \ 
+Y_{e_{\bullet_2}}^2 Y_{e_{\circ_1}}^2 Y_{e_{\circ_2}}^3+Y_{e_{\bullet_1}} Y_{e_{\bullet_2}}^2 Y_{e_{\circ_1}}^2 Y_{e_{\circ_2}}^3 \Big) \, , \cr
& \ \vdots \ &
\end{eqnarray}
In general we generate the whole family with core charges
\begin{equation}
\begin{cases}
& \left( - 2 n e_{\bullet_1} - n e_{\bullet_2} - \frac{6n-2}{3} e_{\circ_1} - \frac{3n-1}{3} e_{\circ_2}  \right)_{n \ge 0} \\
& \left( -n e_{\bullet_1} - 2 n e_{\bullet_2} - \frac13 (3n+1) e_{\circ_1} - \frac{6n+2}{3} e_{\circ_2} \right)_{n \ge 0} 
\end{cases} \, . 
\end{equation}
Similarly starting from $-\frac23 e_{\circ_1} - \frac13 e_{\circ_2}$ we generate line defects with charges
\begin{equation}
\begin{cases}
&  \left( -n e_{\bullet_1} - 2 n e_{\bullet_2} - \frac{3n-1}{3} e_{\circ_1} - \frac{6n-2}{3} e_{\circ_2} \right)_{n \ge 0} \\ 
&  \left( -2n e_{\bullet_1} - n e_{\bullet_2} - \frac23 (3n+1) e_{\circ_1} - \frac{3 n +1}{3} e_{\circ_2} \right)_{n \ge 0}
\end{cases} \, .
\end{equation}
The explicit form of the vevs for these operators is available at \cite{mathe}.

\underline{$\left( \mathrm{SU}(3) / \mathbb{Z}_3 \right)_2$}

This time we add line defecs whose core charges are $\frac13 e_{\bullet_1} + \frac23 e_{\bullet_2}-\frac13 e_{\circ_1}-\frac23 e_{\circ_2}$ and $\frac23 e_{\bullet_1} + \frac13 e_{\bullet_2}-\frac23 e_{\circ_1}-\frac13 e_{\circ_2}$. As before the set of dyonic line defects of $SU(3)$ is unchanged and the Wilson lines which are not invariant under $\mathbb{Z}_3$ are projected out.

From $\frac13 e_{\bullet_1} + \frac23 e_{\bullet_2}-\frac13 e_{\circ_1}-\frac23 e_{\circ_2}$  we generate line defects with core charges
\begin{equation} \label{su3Z32seq1}
\begin{cases}
&  \left( - \frac23 (6n+1) e_{\bullet_1} - \frac{6n+1}{3} e_{\bullet_2} - \frac23 (6n-1) e_{\circ_1} - \frac{6n-1}{3} e_{\circ_2} \right)_{n \ge 0} \\ 
&  \left( - \frac{6n-1}{3} e_{\bullet_1} - \frac23 (6n-1) e_{\bullet_2} - \frac{6n+1}{3} e_{\circ_1}  - \frac23 (6n+1) e_{\circ_2} \right)_{n \ge 0}
\end{cases} \, , 
\end{equation}
while from $\frac23 e_{\bullet_1} + \frac13 e_{\bullet_2}-\frac23 e_{\circ_1}-\frac13 e_{\circ_2}$ we obtain
\begin{equation} \label{su3Z32seq2}
\begin{cases}
&  \left( - \frac{6n+1}{3} e_{\bullet_1} - \frac23 (6n+1) e_{\bullet_2} - \frac{6n-1}{3} e_{\circ_1} - \frac23 (6n-1) e_{\circ_2} \right)_{n \ge 0}  \\
&  \left( -\frac23 (6n-1) e_{\bullet_1} - \frac{6n-1}{3} e_{\bullet_2} - \frac23 (6n+1) e_{\circ_1} - \frac{6n+1}{3} e_{\circ_2} \right)_{n \ge 0}
\end{cases} \, .
\end{equation}
Again the explicit form of the vevs for these operators is available at \cite{mathe}.

Overall the discrete dynamical system give us a glimpse of the structure of the set of line operators for these three theories, $\left( \mathrm{SU}(3) / \mathbb{Z}_3 \right)_n$, with $n=0,1,2$. We find three distinct set which have a non empty intersection, for example on the Wilson lines of $\mathrm{SU}(3)$ which are $\zed_3$-invariant. Each set can be constructed explicitly as the union of the orbits of the dynamical system, each orbit generated starting from a line defect compatible with the rest of the set.

\paragraph{Remark.} We know that the correct operation to generate defects in $\left( \mathrm{SU}(3) / \mathbb{Z}_3 \right)_n$ theories is the half-monodromy of SU(3). But what happens if we insist on using the 1/6 monodromy \eqref{RtopSU3}? This is the IR counterpart of doing a $\theta$-angle rotation with the wrong periodicity. What we expect from the UV analysis of \cite{Aharony:2013hda} is that such operation should map one theory into another. We will show this explicitly with one example, where the action of the 1/6 monodromy maps a defect of one theory into a defect of another theory.

Consider the following sequence of line operators, both their vevs and the associated framed quivers, obtained by iterating the (now unphysical) operation $\mathsf{R}$ derived from \eqref{RtopSU3}:
\begin{align}
\xymatrix@C=6mm{
 f_{-\frac13 e_{\circ_1}-\frac23 e_{\circ_2}}  & \bullet_1 \ar[drr] & & \bullet_2 \ar[dll]  \ar@/_1pc/@{..>}[lll]  &  \langle L \rangle^{\left( SU(3)/\mathbb{Z}_3 \right)_0}  = \frac{1}{Y_{e_{\circ_1}}^{1/3} Y_{e_{\circ_2}}^{2/3}}
  \\
& \circ_1 \ar@<-0.5ex>[u]  \ar@<0.5ex>[u]  & & \circ_2  \ar@<-0.5ex>[u]  \ar@<0.5ex>[u] 
&
\\
\ar@/^/[uu]_{\mathbf{r}^+} f_{\frac13 e_{\bullet_1} + \frac23 e_{\bullet_2}}  &  \bullet_1 \ar[drr] & & \bullet_2 \ar[dll]  &  \langle L \rangle^{\left( SU(3)/\mathbb{Z}_3 \right)_1}  = Y_{e_{\bullet_1}}^{1/3} Y_{e_{\bullet_2}}^{2/3}
 \\
& \circ_1   \ar@<-0.5ex>[u]  \ar@<0.5ex>[u]  & & \circ_2  \ar@<-0.5ex>[u]  \ar@<0.5ex>[u] \ar@/^1pc/@{..>}[ulll]  
&
\\
\ar@/^/[uu]_{\mathbf{r}^+} f_{\frac23 e_{\bullet_1}+\frac13 e_{\bullet_2}+\frac13 e_{\circ_1}+\frac23 e_{\circ_2}} &\bullet_1 \ar[drr] & & \bullet_2 \ar[dll]   \ar@/_1pc/@{<..}[lll] & 
 \langle L \rangle^{\left( SU(3)/\mathbb{Z}_3 \right)_2}  = Y_{e_{\bullet_1}}^{2/3} Y_{e_{\bullet_2}}^{1/3} (1+Y_{e_{\bullet_2}}) Y_{e_{\circ_1}}^{1/3} Y_{e_{\circ_2}}^{2/3}
\\
& \circ_1 \ar@{..>}[ul]  \ar@<-0.5ex>[u]  \ar@<0.5ex>[u]  & & \circ_2  \ar@<-0.5ex>[u]  \ar@<0.5ex>[u] 
&
\\
\ar@/^3pc/[uuuuuu]^{\mathbf{m}^+} \ar@/^/[uu]_{\mathbf{r}^+} f_{\frac23 e_{\circ_1}+ \frac13 e_{\circ_2}} \ar@{..>}[r]  & \bullet_1 \ar[drr] & & \bullet_2 \ar[dll] &  \qquad \langle L \rangle^{\left( SU(3)/\mathbb{Z}_3 \right)_0}  = Y_{e_{\circ_1}}^{2/3} Y_{e_{\circ_2}}^{1/3} \Big( 1+Y_{e_{\bullet_1}}+Y_{e_{\bullet_1}} Y_{e_{\circ_2}}
  \\
& \circ_1 \ar@<-0.5ex>[u]  \ar@<0.5ex>[u]  & & \circ_2  \ar@<-0.5ex>[u]  \ar@<0.5ex>[u] 
 &
 \hspace{3cm} +2 Y_{e_{\bullet_1}} Y_{e_{\bullet_2}} Y_{e_{\circ_2}} +Y_{e_{\bullet_1}} Y_{e_{\bullet_2}}^2 Y_{e_{\circ_2}} \Big)
}
\end{align}
Where the arrows labelled by $\mathbf{r}^+$ and $\mathbf{m}^+$ are not part of the quivers but describe maps between the quivers. Here we have used the $\frac{1}{6}$-fractional monodromy $\mathbf{r}^+$ to connect line defects in different $\left( SU(3) / \mathbb{Z}_3 \right)_n$ theories. Only the half-monodromy $\mathbf{m}^+$, which corresponds to \textsc{cpt} and is given by iterating $\mathbf{r}^+$ three times, maps a theory into itself, $\left( SU(3) / \mathbb{Z}_3 \right)_0$ in the example above. The $\frac{1}{6}$-fractional monodromy has its origin in the $\mathbb{Z}_6$ R-symmetry of $SU(3)$ which is now broken. It corresponds to a shift in the $\theta$-angle of $\theta \longrightarrow \theta + 2 \pi$ which is not anymore a physical shift in $\left( SU(3) / \mathbb{Z}_3 \right)_n$ theories due to the presence of $\frac13 \mathbb{Z}$-charged instantons. 

What we have just described is an IR manifestation of an UV phenomenon noted in \cite{Aharony:2013hda}. In the lagrangian definition of the theory one can imagine shifting the $\theta$-angle by an unphysical quantity. What was noted in \cite{Aharony:2013hda} is that the unphysical shifts map a theory characterized by a set of line operators into a different theory characterized by another set. In the case at hand this can be written as
\begin{equation} \label{UVrelSU3}
\left( SU(3) / \mathbb{Z}_3 \right)_n^{\theta + 2 \pi} = \left( SU(3) / \mathbb{Z}_3 \right)_{(n+1) \ \mathrm{mod} \, 3}^{\theta}
\end{equation}
Remarkable the discrete dynamical systems that we can associate to gauge theories know about the physical and unphysical shifts of the $\theta$-angle in the UV definition of the theory.

Note that the line operator $ \langle L \rangle^{\left( SU(3)/\mathbb{Z}_3 \right)_2}  = Y_{e_{\bullet_1}}^{2/3} Y_{e_{\bullet_2}}^{1/3} (1+Y_{e_{\bullet_2}}) Y_{e_{\circ_1}}^{1/3} Y_{e_{\circ_2}}^{2/3}$ is \textit{not} part of the families (\ref{su3Z32seq1}) or  (\ref{su3Z32seq2}), although it obeys the Dirac quantization condition only with the operators of  $\left( SU(3) / \mathbb{Z}_3 \right)_2$. Indeed these two sequences are only part of the full set of line defects, singled out by the fact that they are particularly simple to obtain. Here we have found another method to compute vevs of families of operators! Indeed they can be inherited from a closely related theory using the unphysical shifts of the $\theta$-angle. Remarkably the relation \eqref{UVrelSU3} between UV labels, translate in the IR in a prescription to compute the vev of new line operators.

\section{The spectrum generator from fixed points} \label{fixedpoints}

In this Section we take an opposite perspective. Assume that by some means we know the functional form of the fixed points of the dynamical system: can we recover the rational transformation? Indeed a point stressed in \cite{Gaiotto:2010be} is that the presence of line defects poses strong constraints on the BPS spectrum. In this Section we will show an explicit realization of this idea, where the knowledge of certain line operators is enough to determine the full spectrum generator. We will begin with an illustrative example in the case of $\mathrm{SU}(2)$ Yang-Mills and then derive explicitly the spectrum generator for the $\mathrm{SU}(2)$ $\cN=2^*$ theory.

\subsection{A simple example: SU(2) Yang-Mills}

We will explain our ideas with a simple example, the case of pure $\rm SU(2)$ super Yang-Mills. In this case both the spectrum generator and the set of line operators are already known explicitly \cite{Gaiotto:2010be}. We choose this example only for illustrative purposes, as the computations can be outlined almost explicitly without too much algebra.

The theory has a basis of charges given by $\{ e_1 , e_2 \}$ with $\langle e_1 , e_2 \rangle = -2$. We know that the fundamental Wilson line is invariant under the action of the half-monodromy \cite{Cirafici:2017iju} as its core charge $-\frac12 (e_1 + e_2)$ is purely electrical and the quantum monodromy acts on the charge via the Witten effect. Such operator was computed in  \cite{Gaiotto:2010be} and can be written as
\begin{equation} \label{SU2Wilson}
W_{\mbf 2} \left[ x,y \right] = \frac{1}{\sqrt{x \, y}} + \sqrt{\frac{y}{x}} + \sqrt{x \, y} \, ,
\end{equation}
with $x= Y_{e_1}$ and $y = Y_{e_2}$. 

Suppose now that the explicit form of the spectrum generator were unknown. We will show how it can be recovered assuming the invariance of the Wilson line \eqref{SU2Wilson}. To simplify the technicalities, let us assume that we know that the model has a fractional monodromy whose permutation exchanges the two nodes. To determine the form of the fractional monodromy we are then led to the ansatz
\begin{equation}
R : (x,  y ) \longrightarrow \left(  y \frac{(c_{00} + c_{10} x + c_{01} y)^2}{(d_{00} + d_{10} x + d_{01} y)^2} , x \frac{(a_{00} + a_{10} x + a_{01} y)^2}{(b_{00} + b_{10} x + b_{01} y)^2} \right) \, .
\end{equation}
In words our ansatz assumes that the action of the fractional monodromy exchanges the two cluster coordinates up to a rational function. This function is squared because $\langle e_1 , e_2 \rangle = -2$. The condition that the Wilson line operator \eqref{SU2Wilson} is invariant under the fractional monodromy transformation is then
\be \label{CondSU2FP}
W_{\mbf 2} \left[ x,y \right]  - W_{\mbf 2} \left[ R (x , y) \right] = 0 \, .
\ee
Of course this equation does not have a uniques solution; assuming we can find such an $R$, any of its iterations will leave the function $W$ invariant. Indeed this is precisely the definition of a fixed point.

Because of the squares, condition \eqref{CondSU2FP} is actually polynomial. We write it as
\be
\frac{ P_6 [x,y]}{\sqrt{y x} \left( 
(a_{00} + a_{10} x + a_{01} y) \, (b_{00} + b_{10} x + b_{01} y) \, (c_{00} + c_{10} x + c_{01} y) \, (d_{00} + d_{10} x + d_{01} y)
\right)}  = 0
\ee
where
\be
P_6 [ x , y] = \sum_{i,j=1}^6 \ m_{ij} x^i y^j
\ee
is a polynomial of degree 6. The coefficients $m_{ij}$ are complicated algebraic functions of the parameters (the interested reader can find them in the supporting \textsc{mathematica} file \cite{mathe}) and must all vanish independently. By direct inspection we see that we can set to zero the coefficients $\{ a_{10} , b_{00} , b_{01} , c_{01} , d_{10} , d_{01} \}$. The system is still overdetermined, so we pick a sample of the conditions and solve them, hoping that the solution will solve also the remaining equations:
\begin{align}
-a_{00} \, b_{10} \, c_{00} \, d_{00} + a_{00}^2 \, d_{00}^2 & = 0 \, ,
\cr
a_{01}^2 \, c_{00}^2 & = 0 \, ,
\cr
-a_{00} \, b_{10} \, c_{10} \, d_{00} + b_{10}^2 \, d_{00}^2 & = 0 \, ,
\cr
a_{00}^2 \, c_{10}^2 - a_{00} \, b_{10} \, c_{10} \, d_{00} & = 0 \, ,
\cr
2 a_{00} \, a_{01} \, c_{10}^2 - a_{01} \, b_{10} \, c_{10} \, d_{00} & = 0 \, .
\end{align} 
This system is not particularly illuminating, but it serve our purposes to illustrate a trick. By experience with cluster transformations we expect all the coefficients to be small natural numbers. We can therefore simply use \textsc{mathematica} to solve this system over the natural numbers (or even a finite field), which in general will make the computation rather fast. This system still does not have a unique solution, but one can easily check that all the solutions obtained as above give rise to the same rational transformation, and also solve the remaining equations. A particularly simple solution sets $a_{01}=0$ and all the remaining parameters equal to one, which gives
\begin{equation}
R \, : \, (x,  y ) \longrightarrow \left( y (1 + x)^2 \ , \ \frac{1}{x}  \right) \, ,
\end{equation}
which agrees with the known results. Indeed iterating this transformation precisely reproduces the half monodromy 
\begin{equation}
R_{1,2} \, : \, (x,  y ) \longrightarrow \left( \frac{\left((x+1)^2 y+1\right)^2}{x},\frac{1}{(x+1)^2 y} \right) \, .
\end{equation}

\paragraph{Remark.} Technically this is an existence result: we have found a particular rational transformation which has the fundamental Wilson line as a fixed point. This transformation is not unique. For example one can check that also the transformation
\begin{equation}
R \, : \, (x,  y ) \longrightarrow \left( \frac{1}{y},\frac{x y^2}{(y+1)^2} \right)
\end{equation}
fixes the fundamental Wilson line. This is indeed the transformation $\tilde{R}$ which arises from the other set of cluster transformations. The two are related by opposing the quiver, and since the BPS quiver for SU(2) is very symmetric, it is expected that the two transformations enjoy similar properties.

We expect that the transformation which we find with this method is ``unique'' up to symmetry transformations of the quiver and up to iterations of smaller monodromies. Furthermore, in case we are in doubt, we can always check that our transformation has the expected properties on the infinite set of all line operators. It would be nice to have a more conclusive argument, but we will leave it for future work.

\subsection{The spectrum generator for SU(2) $\cN=2^*$}

Now we apply our arguments to the case of the SU(2) $\cN=2^*$ theory. In this case the quiver formalism gives partial results on the spectrum, available at certain points in the moduli space  \cite{Cecotti:2015qha}. On the other hand using the formalism of spectral networks one is able to write down a rational transformation by associating certain rational functions to the combinatorial objects which make up the network \cite{Longhi:2015ivt,Longhi:2016wtv}. This transformation coincides with the Kontsevich-Soibelman transformation. We will now show that it is possible to recover this transformation also from the perspective of quivers, by purely algebraic computations, within our formalism.

The BPS quiver is constructed from three nodes $\{ \bullet_1, \bullet_2 , \bullet_3 \}$ corresponding to the basis $\{ e_i \}_{i=1}^3$ with pairing $\langle e_i , e_{i+1} \rangle = +2$ cyclically:
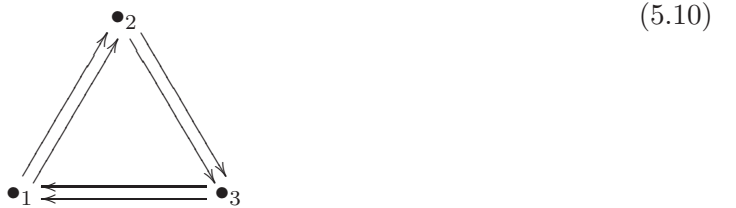
\begin{equation}
\xymatrix@C=8mm{
& & \bullet_2 \ar@<0.5ex>[rdd] \ar@<-0.5ex>[rdd] & \\
& & & \\
& \bullet_1 \ar@<0.5ex>[ruu] \ar@<-0.5ex>[ruu]  & & \bullet_3 \ar@<0.5ex>[ll] \ar@<-0.5ex>[ll] \\
}
\end{equation}
and is also called Markov quiver.
In this case we have three fundamental line operators, which all have the same form \cite{Gaiotto:2010be}. They are a Wilson line, a 't Hooft line and a dyonic line. Their core charge is given by $-\frac12 \left( e_{i} + e_{i+1} \right)$, cyclically. The framed quiver for each of these operators is obtained by adding a closed loop to each one of the Kronecker subquivers of the BPS quiver. Their explicit form of the operators was computed in \cite{Gaiotto:2010be} and is given by
\begin{align}
W_{\mathbf{2}} [x_1,x_2] = \frac{1}{\sqrt{x_1 \, x_2}} + \sqrt{\frac{x_1}{x_2}} + \sqrt{x_1 \, x_2}  \, ,
\end{align}
where $(x_1 , x_2)$ are cyclic permutations of $(x,y,z)$ in this order, and we have set $Y_{\bullet_1} = x$, $Y_{\bullet_2} = y$ and $Y_{\bullet_3} = z$. By partial abuse of notation we use the same symbols as for the SU(2) case.

We will assume that these operators are invariant under the action of the (unknown) half-monodromy, eventually up to a relabelling of the nodes. We can argue for this by considering the Wilson line, which corresponds to the A-cycle of the once-punctured torus in the class $S$ description. At the level of the charges the quantum monodromy acts as the Witten effect \cite{Cirafici:2017iju}, shifting the electric charge of a dyon by its magnetic charge. Wilson lines are precisely the objects which are left invariant, and indeed this argument was used in \cite{Cirafici:2017iju} to identify them with conserved charges of an integrable system. In our case, due to the symmetry of the Markov quiver, we can argue that the same is true for all the aforementioned operators. This is a consequence of the S-duality inherited by the $\cN=2^*$ theory from $\cN=4$ super Yang-Mills.

From the cyclic symmetry of the quiver, it is natural to propose an ansatz for the rational transformation of the form 
\be \label{RopSU2star}
\mathbf{R} : (x,y,z) \longrightarrow \left( 
x \ \frac{P_n (x,y,z)^2}{P_d (x,y,z)^2} \, ,
y \  \frac{P_n (y,z,x)^2}{P_d (y,z,x)^2} \, ,
z \  \frac{P_n (z,y,x)^2}{P_d (z,y,x)^2} \, 
\right) \, .
\ee
We have introduced the two polynomials
\begin{align}
P_n (x,y,z) = \sum_{i,j,k}^2 a_{ijk} x^i \, y^j \, z^k \, , \qquad
P_d (x,y,z) = \sum_{i,j,k}^2 b_{ijk} x^i \, y^j \, z^k \, .
\end{align}

Now we have to impose that this transformation leaves invariant the aforementioned operators, up to permutations. 
We therefore postulate
\be \label{W2condSU2star}
W_{\mathbf{2}} \left[y \  \frac{P_n (y,z,x)^2}{P_d (y,z,x)^2} ,  x \ \frac{P_n (x,y,z)^2}{P_d (x,y,z)^2}   \right] = W_{\mathbf{2}} [x , y] 
\ee
and cyclic permutations of the LHS and of the RHS. More precisely such a transformation does not leave the Wilson lines invariant, but also requires a non-cyclic permutation of the charges.

This can be justified heuristically as follows. Consider for example the Wilson line, attached to the nodes $\bullet_1$ and $\bullet_2$. Then one can take a decoupling limit in which we decouple $\bullet_3$ and recover pure SU(2) Yang-Mills (recall that the Markov quiver can be thought of as arising from the quiver of pure Yang-Mills with $N_f = 2$ by joining the two nodes representing the hypermultiplets; these are the states we are decoupling). Our condition \eqref{W2condSU2star} assumes that the rational transformation \eqref{RopSU2star} has the same properties as the fractional monodromy of pure SU(2), although here we incorporate the permutation in the action on the line operators and not on the definition of  \eqref{RopSU2star}. Note that if this assumption is wrong, the correct transformation will differ from ours by a relabelling of the charges.

Similarly as in the pure SU(2) case, the conditions \eqref{W2condSU2star} can be rewritten as
\begin{align} \label{condW2N2star}
\Big( &-P_d^2 (x,y,z)  P_d^2 (y,z,x)  + P_d (x,y,z) P_n (x,y,z)  P_d (y,z,x)   P_n (y,z,x)  
\cr
 & + x P_d (x,y,z) P_n (x,y,z)  P_d (y,z,x)   P_n (y,z,x)  
+ x P_d (x,y,z) P_n (x,y,z) y  P_d (y,z,x)   P_n (y,z,x) 
\cr & - P_d^2 (x,y,z) y  P_n^2 (y,z,x)  - x P_n^2 (x,y,z) y  P_n^2 (y,z,x) \Big) = 0
\end{align}
and cyclic permutations, provided the quantity (which appears in the denominator of \eqref{W2condSU2star})
\be
\sqrt{x} \, P_d (x,y,z) \, P_n (x,y,z) \, \sqrt{y} \, P_d (y,z,x) \, P_n (y,z,x)
\ee
and cyclic permutations thereof, is non-vanishing.

The condition \eqref{condW2N2star} and its permutations have to hold for any value of the variables $(x,y,z)$. In other words the coefficients of the monomials in $x^i \, y^j \, z^k$ have to vanish separately. This implies a rather large collection of algebraic equations which can be solved using \textsc{mathematica} \cite{mathe}. The result is
\begin{align} \label{resPoly}
P_n (x,y,z) = 1 + y + 2 y z + y z^2 + 2 x y z^2 + x^2 y z^2 + x^2 y^2 z^2 \, ,\cr
P_d (x,y,z) = 1 + z + 2 x z + x^2 z + 2 x^2 y z + x^2 y^2 z + x^2 y^2 z^2 \, .
\end{align}
While the computation to obtain \eqref{resPoly} are somewhat involved, it is not difficult to check that the resulting rational transformation \eqref{RopSU2star} satisfy the conditions \eqref{W2condSU2star}.
 
The transformation   \eqref{RopSU2star} corresponds to the spectrum generator for SU(2) $\cN=2^*$. The expression \eqref{RopSU2star} is at this stage conjectural, since its explicit form depends on certain assumptions, such as the specific action in \eqref{W2condSU2star}. However it agrees with the results of \cite{Longhi:2015ivt,Longhi:2016wtv}.



\section{Line operators and spherical DAHA} \label{sDAHA}

In this Section we will show that certain line operators form a presentation of a spherical double affine Hecke algebra (DAHA) by generators and relations. This result is expected from the work of Oblomkov \cite{Oblomkov} who showed that the deformation quantization of the Hitchin moduli space, in complex structure $J$, gives rise to the sperical DAHA for $\mathfrak{gl} (n)$ in the case of $\cN=2^*$ supersymmetric theory. In this case the Hitchin moduli space can be understood as the moduli space of flat $\mathrm{GL} (n , \complex)$ connections on the once punctured torus. From the point of view of class $S$ theories, Oblomkov's result is therefore an UV result. In this Section we establish the IR counterpart, in the case of the once punctured torus and of the sphere with four punctures. From the IR perspective the spherical DAHA arises as the algebra of IR quantum line operators and is therefore determined by the framed BPS spectrum. We will only consider the rank 2 case but our results are expected to extend to higher rank general field theories. Recent works discussing the relation between line operators and DAHA are \cite{Nawata,Okuda:2019emk}. Furthermore, it was noted in \cite{Razamat:2013jxa} that certain generators of the DAHA act non trivially on the lens space index. For the algebraic definitions we follow \cite{DAHAbook}.

\subsection{The SU(2) $\cN=2^*$ theory and $A_1$ spherical DAHA}

In this case the results of  \cite{Oblomkov} imply a UV relation between field theory and the spherical DAHA.  Here we provide the IR counterpart of this result by giving an explicit construction of the spherical DAHA using framed BPS states, in terms of quantum cluster variables. We show by a direct computation that certain IR line operators give a presentation by generators and relations of the $A_1$ spherical DAHA.

The double affine Hecke algebra $\Hecke[A_1]$ is generated by variables $T$, $X$ and $Y$ with relations
\begin{align}
(T - t) (T + t^{-1}) &= 0 \, ,\cr
T X T &= X^{-1} \, ,\cr
T Y^{-1} T &= Y \, ,\cr
Y^{-1} X^{-1} Y X &= q^{-1} T^{-2} \, ,
\end{align}
and depends on the two parameters $t$ and $q$. 

The spherical subalgebra is defined as $\mathcal{S} \Hecke [A_1] := \mathbf{e} \Hecke [A_1] \mathbf{e}$ in term of the idempotent
\be
\mathbf{e} = \frac{T + t^{-1}}{t+t^{-1}} \, .
\ee
We can give a more explicit description of the spherical subalgebra as follows. Introduce generators
\begin{align}
x &= (X + X^{-1}) \mathbf{e} \, ,\cr
y &= (Y + Y^{-1}) \mathbf{e} \, ,\cr
z &= q^{-1/2} (X \, Y \, T^{-2} + X^{-1} \, Y^{-1} ) \mathbf{e} \, .
\end{align}
Then $\mathcal{S} \Hecke [A_1]$ is the algebra generated by $x$, $y$ and $z$ with relations \cite{DAHAbook}
\begin{align} \label{A1sDAHA}
& [x , y]_q = (q - \frac1q) z \, ,\cr
& [y , z]_q = (q - \frac1q) x \, ,\cr
& [z , x]_q = (q - \frac1q) y \, ,\cr
& q x^2 + \frac1q y^2 + q z^2 - q^{\frac12} x y z = (q^\frac 12 + q^{-\frac12})^2 + (t q^{-\frac12} - t^{-1} q^{\frac12})^2 \, ,
\end{align}
where 
\be
[a , b]_q = q^{\frac12} a\, b - q^{-\frac12} b a \, .
\ee
Note that in the limit $q \longrightarrow 1$ the last relation becomes
\be
x^2+y^2+z^2 - x y x = 4 + (t - t^{-1})^2 \, ,
\ee
which defines the character variety associated to the torus with one puncture. 

In the original formulation by Oblomkov the three generators $x$, $y$ and $z$ are associated with certain holonomies of flat connections on the once punctured torus. In physical language they correspond to the Wilson line operator, the t' Hooft line operators and the Wilson-t' Hooft dyonic operator with charge $(1,1)$. We will now show that we can provide an alternative description of the spherical DAHA by using the IR description of these operators, computed by summing over framed BPS states.

The vacuum expectation value of these operators was computed already in \cite{Gaiotto:2010be} from the study of laminations. In order to describe a noncommutative algebra in the IR we need the quantum version of these operators. However, since in \cite{Gaiotto:2010be} it was shown that all the framed BPS states involved are hypermultiplets, it is immediate to write down the quantum line operators since for hypermultipets the refined Donaldson-Thomas invariant is simply 1. 


The three operators have the form
\begin{align}
W_1 & = \sfX_{\frac{1}{2}(-e_1-e_2)}+
\sfX_{\frac12 (e_1-e_2)}+
\sfX_{\frac12 (e_1+e_2)}
\, , \\
W_2 & = \sfX_{\frac{1}{2}(-e_2-e_3)}+
\sfX_{\frac12 (e_2-e_3)}+
\sfX_{\frac12 (e_2+e_3)}
\, , \\
W_3 & = \sfX_{\frac{1}{2}(-e_1-e_3)}+
\sfX_{\frac12 (e_3-e_1)}+
\sfX_{\frac12 (e_1+e_3)} \, ,
\end{align}
where $\{ e_1 , e_2 , e_3 \}$ is the basis of charges corresponding to the Markov quiver.

We now claim the identification
\be
W_1 \longleftrightarrow x \, , \qquad
W_2 \longleftrightarrow y \, , \qquad
W_3 \longleftrightarrow z \, ,
\ee
while the role of the parameter is played by $t = - \ii q^{-\frac12} \sfX_{\frac12 (e_1+e_2+e_3) }$. We will now show that these line operators indeed form a presentation of $\mathcal{S} \Hecke [A_1]$. Once the relevant generators and parameters are identified, this is done by a direct computation.

To begin with, we discuss the commutation relations. Let us start by evaluating
\begin{align}
q^{\frac12} W_1 W_2 & = q^{\frac12} \left(  \sfX_{\frac{1}{2}(-e_1-e_2)}+
\sfX_{\frac12 (e_1-e_2)}+
\sfX_{\frac12 (e_1+e_2)}
\right)
\left(
 \sfX_{\frac{1}{2}(-e_2-e_3)}+
\sfX_{\frac12 (e_2-e_3)}+
\sfX_{\frac12 (e_2+e_3)}
\right)
\cr
& = 
\sfX_{-\frac{e_1}{2}-e_2-\frac{e_3}{2}}+
\sfX_{\frac{e_1}{2}-e_2-\frac{e_3}{2}}+
\sfX_{\frac{e_1}{2}+e_2-\frac{e_3}{2}}+
\sfX_{\frac{e_1}{2}+e_2+\frac{e_3}{2}}+q \,
\sfX_{-\frac{e_1}{2}-\frac{e_3}{2}}+q \,
\sfX_{\frac{e_1}{2}-\frac{e_3}{2}}
\cr & \ \ +
\frac{1}{q} X_{\frac{e_1}{2}-\frac{e_3}{2}}+q \,
   X_{\frac{e_3}{2}-\frac{e_1}{2}}+q
   X_{\frac{e_1}{2}+\frac{e_3}{2}} \, .
\end{align}
Similarly
\begin{align}
- q^{- \frac12} W_2 W_1 &= 
-\sfX_{-\frac{e_1}{2}-e_2-\frac{e_3}{2}}-
\sfX_{\frac{e_1}{2}-e_2-\frac{e_3}{2}}-
\sfX_{\frac{e_1}{2}+e_2-\frac{e_3}{2}}-
\sfX_{\frac{e_1}{2}+e_2+\frac{e_3}{2}}-
\frac{1}{q} X_{-\frac{e_1}{2}-\frac{e_3}{2}}-
q \, \sfX_{\frac{e_1}{2}-\frac{e_3}{2}}
\cr & \ \ -
\frac{1}{q} \sfX_{\frac{e_1}{2}-\frac{e_3}{2}}-
\frac{1}{q} \sfX_{\frac{e_3}{2}-\frac{e_1}{2}}-
\frac{1}{q} \sfX_{\frac{e_1}{2}+\frac{e_3}{2}} \, .
\end{align}
From which we see
\be
 [W_1 , W_2]_q = (q - \frac1q) W_3 
\ee
The other commutation relations can be checked similarly, since the line operators differ by a relabeling of the charges. Now we turn to the last relation of \eqref{A1sDAHA}. The right hand side is relatively simple to compute
\be
(q^\frac 12 + q^{-\frac12})^2 + (t q^{-\frac12} - t^{-1} q^{\frac12})^2
 =
 -\sfX_{-e_1-e_2-e_3}-\sfX_{e_1+e_2+e_3}+q+\frac{1}{q}
\ee
The left hand side is slightly more involved. We begin with the quadratic terms. First one can see that
\be
q W_1^2 = q \sfX_{-e_1-e_2}+q \sfX_{e_1-e_2}+q \sfX_{e_1+e_2}+q^2
   \sfX_{e_1}+\sfX_{e_1}+q^2 \sfX_{-e_2}+\sfX_{-e_2}+2 q
\ee
The other quadratic terms only differ by a relabeling of the charges. Finally the cubic term is
\begin{align}
- q^{\frac12} W_1 W_2 W_3 &= 
-\sfX_{-e_1-e_2-e_3}-
\sfX_{e_1+e_2+e_3}-
q \,  \sfX_{-e_1-e_2}-
q \, \sfX_{e_1-e_2}-
q \, \sfX_{e_1+e_2}-
q \, \sfX_{-e_1-e_3}-
q \, \sfX_{e_3-e_1}
\cr & \ \ -
q \, \sfX_{e_1+e_3}-
q^2 \, \sfX_{-e_1}-
q^2 \, \sfX_{e_1}-
\sfX_{-e_1}-
\sfX_{e_1}-
\frac{1}{q} \sfX_{-e_2-e_3}-
\frac{1}{q} \sfX_{e_2-e_3}-
\frac{1}{q} \sfX_{e_2+e_3}
\cr & \ \ -
q^2 \, \sfX_{-e_2}-
\frac{1}{q^2} \sfX_{e_2}-
\sfX_{-e_2}-
\sfX_{e_2}-
q^2 \, \sfX_{e_3}-
\frac{1}{q^2} \sfX_{-e_3}-
\sfX_{-e_3}-
\sfX_{e_3}-
3 \, q-\frac{1}{q}
\end{align}
Now it is easy to see that putting everything together the last relation of \eqref{A1sDAHA} is satisfied.

\textbf{Remark.} It should be stressed that the algebra of line operators is a wall-crossing invariant. In particular, if we have found a presentation by generators and relations which holds in one chamber, we are now able to generate many other presentation by framed wall-crossing, that is mutating the quiver.  

\subsection{The SU(2) $N_f=4$ theory and $C^\vee C_1$ spherical DAHA}

In this case the moduli space of vacua of the theory on $\real^3 \times S^1$, the Hitchin moduli space, can be identified with the character variety of a sphere with four punctures. As in the previous case, such a character variety is related to a DAHA, namely the $C^\vee C_1$ spherical subalgebra \cite{Oblomkov2,Ter13}. The character variety is is the space of homomorphisms between the fundamental group and $GL (2,\complex)$, modulo gauge transformations. Of the generators of the fundamental group, those comprising two punctures become generators of the spherical DAHA, upon deformation quantization of the character variety \cite{Oblomkov2,Ter13}.

In the classification of line operators of \cite{Gaiotto:2010be}, such loops correspond precisely to the Wilson loop, the 't Hooft loop and a dyonic loop operators, as in the $\cN=2^*$ case. In the following we will exhibit explicitly these operators in terms of framed quivers. Such framed quivers are invariant under the action of the dynamical system on the quiver. Therefore the associated IR line operator vacuum expectation value is a fixed point of the dynamical system. It is therefore natural to conjecture that the generators of the spherical DAHA should be among the fixed points of the dynamical system. We will now prove this statement by a direct computation. We will use our dynamical system to determine the form of these operators.

The $C^\vee C_1$ double affine Hecke algebra $\Hecke [C^\vee C_1]$ is generated by the elements $T_i$, $i=1,\dots,4$ subject to the relations
\begin{align}
(T_i - t_i) (T_i + t_i^{-1}) = 0, \qquad i=1,2,3,4  \ , \ \qquad T_4 \, T_3 \, T_2 \, T_1 = q \, .
\end{align}
and depends explicitly on the parameters $q \in \complex^*$ and $t_i \in \complex$, $i=1,\dots,4$.

The spherical subalgebra is defined as $\mathcal{S} \Hecke [C^\vee C_1] := \mathbf{e} \Hecke [C^\vee C_1] \mathbf{e}$ where
\be
\mathbf{e} = \frac{T_3 + t_3^{-1}}{t_3 + t_3^{-1}} 
\ee
is an idempotent. A presentation of this spherical subalgebra by generators and relations is as follows. Define the three generators
\begin{align}
x_1 &= (T_4 \, T_3 + (T_4  \,T_3)^{-1}) \mathbf{e} \, , \\
x_2 &= (T_3 \, T_2 + (T_3  \,T_2)^{-1}) \mathbf{e} \, , \\
x_3 &= (T_3 \, T_1 + (T_3  \,T_1)^{-1}) \mathbf{e} \, .
\end{align}
Then  $\mathcal{S} \Hecke [C^\vee C_1]$ is generated by $x_1$, $x_2$ and $x_3$ with the relations (note the different power of $q$ from the  $A_1$ case discussed previously; we use the same letter to avoid unnecessary notation, hoping that this will not cause confusion)
\begin{align}
& \left[ x_1 , x_2 \right]_{q^2} = (q^2 - q^{-2}) x_3 - (q - q^{-1}) \alpha_3 \label{dahacomm1} \, , \\
& \left[ x_2 , x_3 \right]_{q^2} = (q^2 - q^{-2}) x_1 - (q - q^{-1}) \alpha_1 \, , \\
& \left[ x_3 , x_1 \right]_{q^2} = (q^2 - q^{-2}) x_2 - (q - q^{-1}) \alpha_2 \, , \\
& q^2 x_1^2 + q^{-2} \, x_2^2 + q^2 \, x_3^2  - q \, x_1 \, x_2 \, x_3 - q \, \alpha_1 \, x_1 - q^{-1} \, \alpha_2 \, x_2 - q \,\alpha_3 \, x_3  \label{dahachar}  \\
& \qquad \qquad \! \! = \overline{t}_1^2 + \overline{t}_2^2 + (\overline{q \, t_3})^2 + \overline{t_4}^2 - \overline{t}_1 \overline{t}_2 (\overline{q \, t_3} ) \overline{t}_4 + (q + q^{-1})^2 \, , \nonumber
\end{align}
in terms of the parameters
\begin{align} \label{CCDAHAparameters}
& \overline{t}_i = t_i - t_i^{-1} \,, \qquad  i=1,2,4 \, , \qquad \overline{q \, t_3} = q \, t_3 - (q \, t_3)^{-1} \, ,\\
& \alpha_1 = \overline{t}_1 \, \overline{t}_2 + (\overline{q \, t_3}) \overline{t}_4 \, ,
\qquad
\alpha_2 = \overline{t}_1 \, \overline{t}_4 + (\overline{q \, t_3}) \overline{t}_2 \, ,
\qquad
\alpha_3 = \overline{t}_2 \, \overline{t}_4 + (\overline{q \, t_3}) \overline{t}_1  \, .
\end{align}

Now we will show by direct computation that this algebra is realized in terms of the physical quantities of the superconformal SU(2) theory with four flavours.

The theory can be described by the BPS quiver:
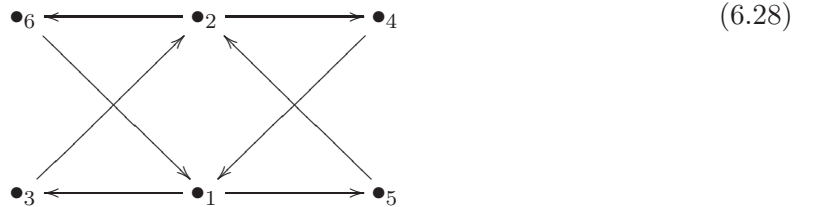
\begin{equation}
\xymatrix@C=8mm{
& \bullet_6 \ar[rrdd] & & \ar[ll] \bullet_2 \ar[rr] & &\ar[lldd]  \bullet_4 \\
& & & & & \\
& \bullet_3  \ar[rruu] & & \ar[ll] \bullet_1 \ar[rr] & & \ar[lluu]  \bullet_5 \\
}
\end{equation}
We can choose a chamber with a $1/4$ monodromy, corresponding to a sequence of mutation on the nodes $\{3,4,5,6, 1, 2 \}$. The full $1/2$ monodromy takes the basis $\{ e_i \}$ into minus itself, up to the permutations $e_3 \leftrightarrow e_5$ and $e_4 \leftrightarrow e_6$.

Consider now the three core charges $\gamma_1 =  \frac{1}{2} (-e_1-e_2-e_4-e_6)$, $\gamma_2 = \frac{1}{2} (-e_1-e_2-e_3-e_5)$ and $\gamma_3 = \frac{1}{2} (-2e_2-e_3-e_4-e_5-e_6)$. One can verify directly that the corresponding framed quivers are left invariant by the sequence of quiver mutations corresponding to the 1/4-monodromy, up to the permutation of the nodes $(1,3,5) \leftrightarrow (2,4,6)$. 

According to our formalism it is natural to conjecture that the line operators we seek are among the fixed points of the associated quantum dynamical system. Indeed, as in the previous case, one of these operators should correspond in the UV to the holonomy of a purely electrical quantum state, and the others related to it by symmetries. We will now confirm this expectation by a direct computation.

We claim that the fixed point conditions are solved by the three operators
\begin{align}
W_1 & = \sfX_{\frac{1}{2} (-e_1-e_2-e_4-e_6)}+
\sfX_{\frac{1}{2}(-e_1+e_2-e_4-e_6)}+
\sfX_{\frac{1}{2}(-e_1+e_2+e_4-e_6)}
\cr & \ +
\sfX_{\frac{1}{2}(-e_1+e_2-e_4+e_6)}+
\sfX_{\frac{1}{2}(-e_1+e_2+e_4+e_6)}+
\sfX_{\frac{1}{2}(e_1+e_2+e_4+e_6)}
\\
W_2 & = \sfX_{\frac{1}{2} (-e_1-e_2-e_3-e_5)}+
\sfX_{\frac{1}{2}(e_1-e_2-e_3-e_5)}+
\sfX_{\frac{1}{2}(e_1-e_2+e_3-e_5)}
\cr & \ +
\sfX_{\frac{1}{2}(e_1-e_2-e_3+e_5)}+
\sfX_{\frac{1}{2}(e_1-e_2+e_3+e_5)}+
\sfX_{\frac{1}{2}(e_1+e_2+e_3+e_5)}
\\
W_3 &=
\sfX_{\frac{1}{2} (-2e_2-e_3-e_4-e_5-e_6)}+
\sfX_{\frac{1}{2} (-2e_2+e_3-e_4-e_5-e_6)}+
\sfX_{\frac{1}{2} (-2e_2-e_3-e_4+e_5-e_6)}+
\sfX_{\frac{1}{2} (-2e_2+e_3-e_4+e_5-e_6)}
\cr & \ +
\sfX_{\frac{1}{2} (2e_2+e_3-e_4+e_5-e_6)}+
\sfX_{\frac{1}{2} (2e_2+e_3+e_4+e_5-e_6)}+
\sfX_{\frac{1}{2} (2e_2+e_3-e_4+e_5+e_6)}+
\sfX_{\frac{1}{2} (2e_2+e_3+e_4+e_5+e_6)}
\cr & \ +
\left(q+\frac{1}{q}\right)  \sfX_{\frac{1}{2} (e_3-e_4+e_5-e_6)}+
\sfX_{\frac{1}{2}(e_3-e_4-e_5-e_6)}+
\sfX_{\frac{1}{2}(-e_3-e_4+e_5-e_6)}+
\sfX_{\frac{1}{2}(e_3+e_4+e_5-e_6)}
\cr & \ +
\sfX_{\frac{1}{2}(e_3-e_4+e_5+e_6)}
\end{align}


The computation to derive such operators are again somewhat involved. However once they are found it is easier to check that they satisfy the requested properties. Consider the first operator. Since all the framed BPS states have framed degeneracy equal to one, it is equivalent to check the invariance of such operator in the classical $q \longrightarrow 1$ limit. In this limit we denote  the coordinates by $Y_{\bullet_i} = y_i$. Then the rational transformation corresponding to the fractional monodromy is obtained by the sequence of mutations $\{ 3, 4, 5, 6, 1, 2 \}$ and is given by
\begin{align}
y_1 &\longrightarrow \frac{(y_4+1) (y_6+1)}{y_1 (y_3+1) y_4 (y_5+1)
   y_6} \, , \\
y_3 &\longrightarrow \frac{y_2 y_5 (y_4 (y_6 (y_1 (y_3+1)
   (y_5+1)+1)+1)+y_6+1)}{y_3 (y_5 (y_2 (y_4+1)
   (y_6+1)+1)+1)+y_5+1}
   \, , \\
y_5 &\longrightarrow\frac{y_2 y_3 (y_4 (y_6 (y_1 (y_3+1)
   (y_5+1)+1)+1)+y_6+1)}{y_3 (y_5 (y_2 (y_4+1)
   (y_6+1)+1)+1)+y_5+1} \, , 
\end{align}
and the transformation laws for $y_2$,$y_4$ and $y_6$ are the same as for $y_1$,$y_3$ and $y_5$ respectively with the permutation of the labels $(1,3,5) \leftrightarrow (2,4,6)$. It is now only a matter of brute force to check that $W_1 [R \{ y_1 , y_2 , y_3 , y_4 , y_5, y_6 \}] =W_1 [\{ y_2 , y_1 , y_4 , y_3 , y_6, y_5 \}]  $ (recall that the fractional monodromy leaves invariant the framed quiver up to the permutation of the nodes  $(1,3,5) \leftrightarrow (2,4,6)$). Once we have shown that the classical operator is invariant, since all the framed BPS states correspond to hypermultiplets, we derive immediately the form of the quantum operator since the refined BPS degeneracy of an hypermultiplet is one as well.

Similar arguments hold for $W_2$. The situation for $W_3$ is slightly more complicated, because there is a framed BPS state with nontrivial spin. However we can still check the invariance of the operator in the commutative limit, where the $q  + 1/q$ factor is replaced by a two. Once the classical operator is checked to be invariant, one could rightly object that a coefficient equal to $2$ could represent a vector multiplet but also two hypermultiplets and the lift to the quantum operator is ambiguous. To solve this ambiguity and show that indeed there is a factor $q  + 1/q$ one can for example do a direct localization computation using the formalism of \cite{Cirafici:2019otj}, which indeed confirms the presence of a vector multiplet. Equivalently one could check the invariance of the operator using the quantum fractional monodromy, but this is more involved.

We now claim the identification
\be
W_1 \longleftrightarrow x_1 \, , \qquad
W_2 \longleftrightarrow x_2 \, , \qquad
W_3 \longleftrightarrow x_3 \, ,
\ee
between the physical quantum line operators and the generators of $\mathcal{S} \Hecke [C^\vee C_1]$. To show this, we begin by introducing the parameters, as in \eqref{CCDAHAparameters}
\begin{align}
\overline{t}_1 &= - \ii \left( \sfX_{- \frac12 (e_1 + e_2 + e_3 + e_6)} + \sfX_{\frac12 (e_1 + e_2 + e_3 + e_6)} \right) \, , \\
\overline{t}_2 &= - \ii \left( \sfX_{ - \frac12 (e_3 + e_4)} + \sfX_{ \frac12 (e_3 + e_4)} \right) \, , \\
\overline{q \, t_3} &= - \ii \left( \sfX_{ - \frac12 (e_5 + e_6)} + \sfX_{\frac12 (e_5 + e_6)} \right) \, , \\
\overline{t}_4 &= -\ii \left( \sfX_{- \frac12 (e_1 + e_2 + e_4 + e_5)} + \sfX_{\frac12 (e_1 + e_2 + e_4 + e_5)} \right) \, .
\end{align}
Note that all the relevant expressions are quadratic in these quantities. The charges involved are the simplest linear combinations which have vanishing pairing with each element of the basis $\{e_i \}$. Because of this the quantum coordinates commute and it is immediate to see that the remaining parameters in \eqref{CCDAHAparameters} are given by
\begin{align}
\alpha_1 & = 
-\sfX_{-\frac{e_1}{2}-\frac{e_2}{2}-e_3-\frac{e_4}{2}-\frac{e_6}{2}}-
\sfX_{\frac{e_1}{2}+\frac{e_2}{2}+e_3+\frac{e_4}{2}+\frac{e_6}{2}}-
   \sfX_{-\frac{e_1}{2}-\frac{e_2}{2}-\frac{e_4}{2}-e_5-\frac{e_6}{2}}-
   \sfX_{\frac{e_1}{2}+\frac{e_2}{2}+\frac{e_4}{2}+e_5+\frac{e_6}{2}} \cr & -
   \sfX_{-\frac{e_1}{2}-\frac{e_2}{2}+\frac{e_4}{2}-\frac{e_6}{2}}-
   \sfX_{\frac{e_1}{2}+\frac{e_2}{2}+\frac{e_4}{2}-\frac{e_6}{2}}-
   \sfX_{-\frac{e_1}{2}-\frac{e_2}{2}-\frac{e_4}{2}+\frac{e_6}{2}}-
   \sfX_{\frac{e_1}{2}+\frac{e_2}{2}-\frac{e_4}{2}+\frac{e_6}{2}}
\, , \\   
\alpha_2 & = 
-\sfX_{-\frac{e_1}{2}-\frac{e_2}{2}-\frac{e_3}{2}-e_4-\frac{e_5}{2}}-
   \sfX_{\frac{e_1}{2}+\frac{e_2}{2}+\frac{e_3}{2}+e_4+\frac{e_5}{2}}-
   \sfX_{-\frac{e_1}{2}-\frac{e_2}{2}-\frac{e_3}{2}-\frac{e_5}{2}-e_6}-
   \sfX_{\frac{e_1}{2}+\frac{e_2}{2}+\frac{e_3}{2}+\frac{e_5}{2}+e_6} \cr & -
   \sfX_{-\frac{e_1}{2}-\frac{e_2}{2}+\frac{e_3}{2}-\frac{e_5}{2}} -
   \sfX_{\frac{e_1}{2}+\frac{e_2}{2}+\frac{e_3}{2}-\frac{e_5}{2}}-
   \sfX_{-\frac{e_1}{2}-\frac{e_2}{2}-\frac{e_3}{2}+\frac{e_5}{2}}-
   \sfX_{\frac{e_1}{2}+\frac{e_2}{2}-\frac{e_3}{2}+\frac{e_5}{2}}
 \, , \\
 \alpha_3 & = 
 -\sfX_{-e_1-e_2-\frac{e_3}{2}-\frac{e_4}{2}-\frac{e_5}{2}-\frac{e_6}{2}}-
   \sfX_{e_1+e_2+\frac{e_3}{2}+\frac{e_4}{2}+\frac{e_5}{2}+\frac{e_6}{2}}-
   \sfX_{-\frac{e_3}{2}-\frac{e_4}{2}-\frac{e_5}{2}-\frac{e_6}{2}}-
   \sfX_{\frac{e_3}{2}+\frac{e_4}{2}-\frac{e_5}{2}-\frac{e_6}{2}} \cr & -
   \sfX_{-\frac{e_3}{2}+\frac{e_4}{2}+\frac{e_5}{2}-\frac{e_6}{2}}-
   \sfX_{\frac{e_3}{2}-\frac{e_4}{2}-\frac{e_5}{2}+\frac{e_6}{2}}-
   \sfX_{-\frac{e_3}{2}-\frac{e_4}{2}+\frac{e_5}{2}+\frac{e_6}{2}}-
   \sfX_{\frac{e_3}{2}+\frac{e_4}{2}+\frac{e_5}{2}+\frac{e_6}{2}}
   \, .
\end{align}

Consider the first relation \eqref{dahacomm1}. Then we see that
\begin{align}
& \left[ W_1 , W_2 \right]_q = \left( q^3 - \frac{1}{q^3} \right) 
\sfX_{\frac{1}{2} (e_3-e_4+e_5-e_6)}
\cr & +  \left( q^2 - \frac{1}{q^2} \right) \Big[
\sfX_{\frac{1}{2} (-2 e_2-e_3-e_4-e_5-e_6)}+
   \sfX_{\frac{1}{2} (-2 e_2+e_3-e_4-e_5-e_6)}+
   \sfX_{\frac{1}{2} (-2 e_2-e_3-e_4+e_5-e_6)}  \cr & +
   \sfX_{\frac{1}{2} (-2 e_2+e_3-e_4+e_5-e_6)}+
   \sfX_{\frac{1}{2} (2 e_2+e_3-e_4+e_5-e_6)}+
   \sfX_{\frac{1}{2} (2 e_2+e_3+e_4+e_5-e_6)} +
   \sfX_{\frac{1}{2} (2 e_2+e_3-e_4+e_5+e_6)} \cr & +
   \sfX_{\frac{1}{2} (2 e_2+e_3+e_4+e_5+e_6)}+
   \sfX_{\frac{1}{2} (e_3-e_4-e_5-e_6)} +
   \sfX_{\frac{1}{2} (-e_3-e_4+e_5-e_6)}+
   \sfX_{\frac{1}{2} (e_3+e_4+e_5-e_6)}+
   \sfX_{\frac{1}{2} (e_3-e_4+e_5+e_6)}
\Big]
\cr & +  \left( q - \frac{1}{q} \right) \Big[
\sfX_{\frac{1}{2} (-2 e_1-2 e_2-e_3-e_4-e_5-e_6)}+
\sfX_{\frac{1}{2} (2e_1+2 e_2+e_3+e_4+e_5+e_6)}+
\sfX_{\frac{1}{2} (-e_3-e_4-e_5-e_6)} \cr &  +
\sfX_{\frac{1}{2} (e_3+e_4-e_5-e_6)}+
\sfX_{\frac{1}{2} (e_3-e_4+e_5-e_6)}+
\sfX_{\frac{1}{2} (-e_3+e_4+e_5-e_6)}+
\sfX_{\frac{1}{2} (e_3-e_4-e_5+e_6)}+
\sfX_{\frac{1}{2}(-e_3-e_4+e_5+e_6)} \cr &  +
\sfX_{\frac{1}{2} (e_3+e_4+e_5+e_6)}
\Big]
\cr &
=  (q^2 - q^{-2}) W_3 - (q - q^{-1}) \alpha_3 
\end{align}
Note that the $ \left( q^3 - \frac{1}{q^3} \right) $ descends directly from the appearance of a vector multiplet in $W_3$. Such a state is crucial to close the algebra. The other commutation relations can be checked similarly.

It is somewhat more involved to check \eqref{dahachar}. To begin with, the left hand side of \eqref{dahachar} can be rewritten as
\begin{align}
&\overline{t_1}^2 + \overline{t_2}^2 + (\overline{q \, t_3})^2 + \overline{t_4}^2 - \overline{t_1} \overline{t_2} (\overline{q \, t_3} ) \overline{t}_4 + (q + q^{-1})^2 \cr
&= -
\sfX_{-e_1-e_2-e_3-e_4-e_5-e_6}-
\sfX_{e_1+ e_2+e_3+e_4+e_5+e_6}-
\sfX_{-e_1-e_2-e_3-e_4}-
\sfX_{e_1+e_2+e_3+e_4}-
\sfX_{-e_1-e_2-e_3-e_6}
\cr & \ -
\sfX_{e_1+e_2+e_3+e_6}-
\sfX_{-e_1-e_2-e_4-e_5}-
\sfX_{e_1+e_2+e_4+e_5}-
\sfX_{-e_1-e_2-e_5-e_6}-
\sfX_{e_1+e_2+e_5+e_6}-
\sfX_{-e_1-e_2}
\cr & \ -
\sfX_{e_1+e_2}-
\sfX_{-e_3-e_4}-
\sfX_{e_3+e_4}-
\sfX_{e_3-e_5}-
\sfX_{e_5-e_3}-
\sfX_{-e_3-e_6}-
\sfX_{e_3+e_6}-
\sfX_{-e_4-e_5}-
\sfX_{e_4+e_5} 
\cr & \ -
\sfX_{e_4-e_6}-
\sfX_{e_6-e_4}-
\sfX_{-e_5-e_6}-
\sfX_{e_5+e_6}+\left(q+\frac{1}{q}\right)^2-8 \, .
\end{align}

The right hand side of \eqref{dahachar} is however much more involved, containing tens of terms with delicate cancellations. Still a computation with \textsc{mathematica} which can be found in \cite{mathe} shows that \eqref{dahachar} is satisfied. Therefore  these three line operators of the SU(2) $N_f = 4$ superconformal field theory form a presentation of $\mathcal{S} \Hecke [C^\vee C_1]$ by generators and relations.

\section*{Acknowledgements}

I wish to thank Michele del Zotto and Davide Polini for collaborations on related projects. I am a member of INDAM-GNFM and of IGAP, I am supported by INFN via the Iniziativa Specifica GAST and by the FRA2018 project ``K-theoretic Enumerative Geometry in Mathematical Physics''.

\end{document}